\documentclass[prb,preprint,amsmath,amssymb]{revtex4}
\begin{document}
\author{Kevin Leung}
\affiliation{Sandia National Laboratories, MS 1415,
Albuquerque, NM 87185\\
\tt kleung@sandia.gov}
\date{\today}
\title{First Principles Modeling of the Initial Stages of
Organic Solvent Decomposition on Li$_x$Mn$_2$O$_4$ (100) Surfaces}

\input epsf
%\ssp
 
\begin{abstract}

Density functional theory and {\it ab initio} molecular dynamics
simulations are applied to investigate the initial steps of ethylene
carbonate (EC) decomposition on spinel Li$_{0.6}$Mn$_2$O$_4$ (100) surfaces.
EC is a key component of the electrolyte used in lithium ion batteries.
We predict an slightly exothermic EC bond breaking event on this oxide facet,
which facilitates subsequent EC oxidation and proton
transfer to the oxide surface.  Both the proton and the partially
decomposed EC fragment weaken the Mn-O ionic bonding network.  Implications
for interfacial film made of decomposed electrolyte on cathode surfaces,
and Li$_x$Mn$_2$O$_4$ dissolution during power cycling, are discussed.
\vspace*{0.5in}
\noindent keywords: lithium ion batteries; lithium manganate; ethylene
carbonate; density functional theory; {\it ab initio} molecule dynamics;
computational electrochemistry

\end{abstract}

\maketitle
 
\section{Introduction}

Lithium ion batteries (LIB) featuring transition metal oxide
cathodes and organic solvent-based electrolytes are currently the 
energy storage devices used to power electric vehicles.  The high
operational voltages of LIBs allow them to store significant amount of energy.
However, these voltages can also induce electrochemical side reactions at
electrode-electrolyte interfaces.\cite{book2,book,xu_review} These lead to
irreversible capacity loss, power fade, durability, and safety issues.
Understanding and controlling these interfacial reactions are of great
interest for improving electric vehicles.  

In this work, we focus on the basic science of one such reaction, and
apply first principles computational techniques to study the first
steps of ethylene carbonate (EC) decomposition on 
the (100) surface of spinel Li$_x$Mn$_2$O$_4$ used as the cathode material
in many lithium ion batteries.\cite{thackeray_rev,thackeray,thackeray98,ceder} 
EC (Fig.~\ref{fig1}) is an indispensible electrolyte component for batteries
that rely on graphitic carbon anodes.  At the negative voltages needed to
charge up a battery and intercalate Li$^+$ into graphite, EC-containing
electrolyte decomposes to form stable solid-electrolyte interphase (SEI)
films that prevent further electron leakage to the
electrolyte.\cite{book2,book,xu_review} As such, substantial theoretical
work has been devoted to excess electron-induced EC chemical reactions in the
vicinity of the anode.\cite{bal01,han,vollmer,ec,bal11,brown}   One may
argue that its ubiquity and critical role make EC in LIB the battery
equivalent of the H$_2$O molecule in biology, geochemistry, and many
solid-liquid interfacial science disciplines.  The present work is indeed
modeled after theoretical studies on water-on-mineral
surfaces.\cite{selloni1,selloni2,selloni3,selloni_rev}

Oxidative reactions of organic electrolytes on cathode surfaces and
formation of surface film from electrolyte decomposition products there\cite{joho,novak1,eriksson,eriksson1,matsushita,kanamura,kanamura1,matsuta,aurbach,aurbach1,aurbach2,moshkovich,jang1997,jang1996,yang2010,novak2,novak3,novak4,guyomard,hirayama,hirayama1,pasquier,matsuo,xia,macneil3,amatucci,xu,amine_mno2,ufheil,arakawa,mn_affect_c0,mn_affect_c1,mn_affect_c2,mn_affect_c3,mn_affect_c4}
are arguably less well understood than reduction on anode surfaces, especially
at the atomic lengthscale.  Electrolyte decomposition products such as
LiF,\cite{aurbach} acetone,\cite{ufheil} aldehydes,\cite{moshkovich} carbon
dioxide,\cite{joho,novak1,moshkovich} organic radicals,\cite{matsuta} and
unidentified polymer, polyether, and carboxylic acid
species\cite{eriksson,matsushita,kanamura,kanamura1,yang2010}
have been detected either in gas form or on the surfaces of 
Li$_x$Mn$_2$O$_4$, as well as on Li$_x$CoO$_2$ and noble metal electrodes 
set at voltages comparable to those of Li$_x$Mn$_2$O$_4$.  The salt used
in the electrolyte strongly affects the surface film composition and
properties.\cite{eriksson,eriksson1,jang1997,mn_affect_c0,mn_affect_c2,mn_affect_c3}
For example, LiPF$_6$ yields fluorine compounds that slow down
Li$^+$ transport into the cathode.\cite{aurbach,jang1997}

The electrode is also degraded by side reactions.  After cycling power,
Li$_x$Mn$_2$O$_4$, exhibits surface morphology changes and loss of Mn
and oxygen ions.\cite{thackeray_rev,thackeray,thackeray98,novak3,novak4,hirayama,hirayama1,aurbach,pasquier,xia,jang1997,amatucci,mn_affect_c2,mn_affect_c3}
It has been widely accepted that Mn(III) ions on spinel electrode surfaces
disproportionate into Mn(IV) and Mn(II),\cite{hunter} the latter of which
dissolves into the electrolyte.  The presence of acid has been shown to
facilitate dissolution and cathode degradation.\cite{thackeray}  In organic
solvents, acid has been speculated to come from reactions
between PF$_6^-$ and trace water,\cite{pasquier} or from organic solvent
decomposition products themselves.\cite{jang1997}  Counterions other than
PF$_6^-$ cause slower Mn$_2$O$_4$ dissolution.\cite{pasquier,jang1997}  There
is also evidence that the emergence of proton-intercalated surface
regions\cite{pasquier} or other surface phases,\cite{xia} not just the
loss of Mn active mass from the electrode, is responsible for
cathode capacity loss.  Dissolved Mn ions have been shown to diffuse to
the anode, degrading the SEI there and weakening anode
passivation.\cite{mn_affect_c2,mn_affect_c3,mn_affect_c0,mn_affect_c1,mn_affect_c4}
Thus anode and cathode degradation are closely related.  
Despite these key insights from experimental efforts, the free energy changes,
activation barriers, specific surface configurations, and reaction pathways
associated with Mn(II) dissolution have not been elucidated in organic
electrolytes.

On the computational side, studies have focused either on cathode oxide surfaces
without solvent molecules,\cite{benedek,maphanga,balach,oxford,ouyang2}
or on anions\cite{ue,johansson} or EC$^+$/PC$^+$ (with an electron removed) in
the absence of the oxide.\cite{zhang_pugh,xing1,xing2,borodin,borodin1}  In
studies of oxide-vacuum interfaces,\cite{benedek,maphanga,balach,oxford,ouyang2}
it has been shown that under-coordinated Mn ions on Mn$_2$O$_4$ surfaces
adopt low oxidation states, and these Mn have been argued to be susceptible
to dissolution.\cite{ouyang2}  However, dissolution
likely require that solvent molecules first directly coordinate to
surface Mn ions,\cite{benedek_new} which restores part of the octahedral
coordination environment.  When this happens, it is not obvious that
the surface Mn(II) and Mn(III) ions do not revert back to Mn(IV).  Explicit
inclusion of solvent molecules on electrode surfaces would be extremely useful
to investigate conjectures about surface Mn charge states.\cite{acid_bulk}

The oxidation potentials of EC (and the similar PC)\cite{zhang_pugh} and
electrolyte anions\cite{ue,johansson} such as PF$_6^-$ have been computed
using DFT methods.  These calculations assume a bulk liquid environment
described by a dielectric constant $\epsilon_o$, and an intact EC molecule
(no ring opening reactions during oxidation).  The predicted oxidation
potentials are at least 1.5~V above the maximum voltage of 4.3~V applied in
many electrolyte oxidation experiments for undoped spinel 
Li$_x$Mn$_2$O$_4$,\cite{eriksson,matsuta,moshkovich,jang1997,amine_mno2}
in broad agreement with tabulated data obtained using inert
electrodes.\cite{xu_review}  Recently, intriguing theoretical studies have
predicted that PF$_6^-$ or
ClO$_4^-$ in the vicinity of PC, with $\epsilon_o$ reduced to reflect
``near-electrode'' conditions, lowers the PC oxidation potential and leads
to novel low-barrier reaction pathways.\cite{borodin,borodin1} The anions
act as proton shuttles and/or provide fluorine atoms to react with PC.
The oxidation steps are accompanied by bond-breaking events, which occur
spontaneously during geometry optimization of (say) the [EC/PF$_6^-$]$^+$ pair.
However, the predicted redox potential of this pair (4.94~V) remains higher
than 4.3~V even after the local $\epsilon_o$ is set to unity.\cite{borodin1}
This suggests that the electrode surface plays a role in catalyzing electrolyte
decomposition.  Despite the assumption of the proximity of the cathode,
the electrode itself is not explicitly included in the calculations.
Nevertheless, the predicted reaction mechanisms successfully reproduces
organic products observed in experiments.\cite{borodin}

In the present work, the effect of Li$_{0.6}$Mn$_2$O$_4$ electrodes on EC
oxidative decomposition is explicitly considered at the atomic lengthscale.  The
$x=0.6$ Li content is roughly at the halfway point of charging or discharging.
Our focus is on the initial reactions that modify and coat the oxide surface,
not on the final distribution of gas/organic products.  Our calculations do
not extend to the high voltage region which seems to be required for CO$_2$,
one of the proposed final products.\cite{novak4} Given the inevitable mixture
of crystal facet/termination in battery cathode materials and the myriad
effects of applied voltage, state-of-charge (lithium content), temperature,
electrode cracking, sweep rate, presence of conductive carbon black,
dissolution of surface films formed by electrolyte decomposition, and
interference from salts, contaminants, and even processes at the anode,
adopting all experimental conditions to these theoretical calculations is
extremely difficult.  Instead, we take
a basic science starting point and focus on EC breakdown on defect-free
Li$_{0.6}$Mn$_2$O$_4$ (100) surfaces.  Our predictions can potentially
be verified with experiments conducted under ideal conditions, but
the mechanism predicted and insights obtained may still be relevant
to battery conditions and may motivate new experiments and
interpretations of battery degradation processes.  

Our approach resembles the intensive study of H$_2$O adsorption on rutile
(011),\cite{selloni1} anatase (101),\cite{selloni2,selloni3} anatase
(001),\cite{selloni3} and many other surface facets of technologically
important titanium oxide over more than a decade.  As in the case of
water-oxide interfaces, both ultra-high vacuum (UHV) and liquid-immersed
electrode conditions are considered in our interfacial models.  These
conditions yield qualitatively similar predictions.  Like H$_2$O on some
TiO$_2$ facets, EC is found to chemisorb on Li$_x$Mn$_2$O$_4$ (100), breaking
internal covalent bonds in the process.  This initial decomposition
step enables the subsequent oxidation reaction which involves electron
and proton transfer to the oxide surface.  This suggests that
organic solvent decomposition and Mn$_2$O$_4$ dissolution may be related.
No assumption of a local $\epsilon_o$ is needed in our calculations.

The (100) surface has been the focus of most theoretical studies on spinel
Li$_x$Mn$_2$O$_4$.\cite{benedek,ouyang2,benedek_new}  Although the 
(111) surface is most prominent in LiMn$_2$O$_4$ particles,\cite{huang2001}
(100) facets remain even after annealing at 800~$^o$C for
100~hours.\cite{huang2001}  The (100) surface has also been synthesized
directly on strontium titanate substrates.\cite{hirayama1}  

This work is organized as follows.  Section~2 describes the theoretical
method used.  Section~3 discusses EC decomposition on Li$_{0.6}$Mn$_2$O$_4$.
Further discussions in Sec.~4 put the results on ideal model surfaces
in the context of battery operating conditions and provide more comparison
with the theoretical literature.  Sec.~5 recapitulates the key findings,
and two appendices describe the charge states of the Mn ions and the
reactions of H$_2$O contaminants and hydroxyl groups on
Li$_x$Mn$_2$O$_4$ (100) surfaces.

\section{Methods}
\label{method}

Static (T=0~K) DFT+U and finite temperature {\it ab initio} molecular dynamics
(AIMD)\cite{aimd} simulations are conducted under UHV and solvent-immersed
electrode conditions, respectively.

The majority of the calculations are performed using the Vienna Atomic
Simulation Package (VASP) version 4.6\cite{vasp,vasp1} and the PBE
functional.\cite{pbe} Modeling spinel Li$_x$Mn$_2$O$_4$ requires the DFT+U
augmented treatment\cite{dftu1} of Mn $3d$ orbitals.  The $U$ and $J$ values
depend on the orbital projection scheme and DFT+U implementation details;
here $U-J=$4.85~eV is chosen in accordance with the literature.\cite{zhou}
A few PBE0 hybrid DFT functional\cite{pbe0} calculations are also performed
using VASP version 5.2.\cite{vasp2}  This more costly method is meant to
provide spot checks of DFT+U results.  A 400~eV planewave energy cutoff is
imposed in all cases.

Static geometry optimization and climbing image nudged elastic band
(NEB)\cite{neb} barrier calculations are performed using a 10$^{-4}$~eV
convergence criterion and 1$\times$2$\times$2~Brillouin zone sampling.
The simulation cells are of 27.59$\times$8.40$\times$8.40$\times$\AA$^3$
dimensions and have a Li$_6$Mn$_{20}$O$_{40}$ stochiometry, not counting
the EC molecule.  The bare lithiated spinel slab is generated as follows.
A (100) Li$_{10}$Mn$_{20}$O$_{40}$ surface slab is relaxed to
its optimal geometry.  The predicted 0.69~J/m$^2$ surface energy agree with
the literature value.\cite{benedek} To reach the target stochiometry, 4~Li
atoms need to be removed.  There are 210~possibilities, many of which are
related by symmetry.  The lowest energy configuration
(Fig.~\ref{fig2}) is found to be the one where a Li on each surface and
half (two) of the Li atoms in the middle layer of the slab are removed.
The lattice constants are kept at those of LiMn$_2$O$_4$.  A larger
system, Li$_8$Mn$_{24}$O$_{48}$, is also discussed in appendix~A.

AIMD simulations feature 40$\times$11.88$\times$11.88\AA$^3$ simulation
cells with a Li$_{12}$Mn$_{40}$O$_{80}$ oxide slab generated by expanded
the oxide slab used in static calculations twofold.  32~EC molecules are
confined between the two surfaces of the slab in the periodically replicated
simulation cell.  $\Gamma$-point Brillouin zone sampling and a
10$^{-6}$~eV convergence criterion are applied at each Born-Oppenheimer
time step.  The trajectories are kept at an average temperature of
T=450~K using Nose thermostats.  Tritium masses on EC are substituted
for protons to permit a time step of 1~fs.  Under these conditions,
the trajectories exhibit drifts of less than 2~K/ps.  EC configurations
are pre-equilibrated using Monte Carlo simulations and simple molecular
force fields, as described in an earlier work.\cite{ec}  We also report
AIMD simulations of hydroxylated MnO$_2$ surfaces in appendix~B, using
deuterium masses for H~atoms and a 0.5~fs time step there.

\begin{figure}
%\centerline{\hbox{ \epsfxsize=3.00in \epsfbox{fig1a.ps} }}
\centerline{\hbox{ \epsfxsize=4.50in \epsfbox{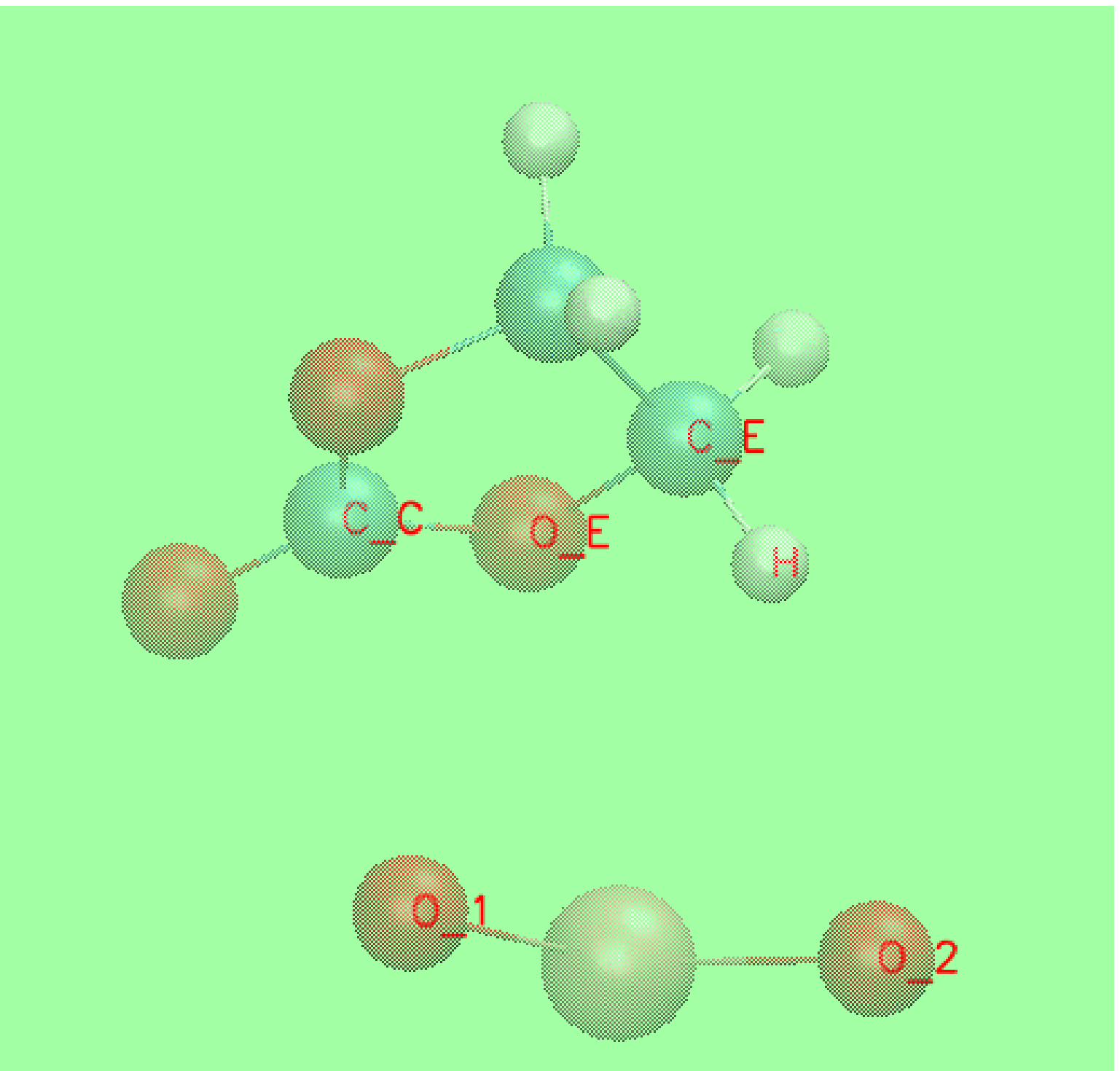} }}
\caption[]
{\label{fig1} \noindent
Ethylene carbonate (EC) molecule.
The panel depicts an isolated EC weakly bound to the surface; three
Li$_x$Mn$_2$O$_4$ atoms are also depicted to explain the labeling
scheme used in the text.  Purple, dark blue, red, light blue, and
white spheres represent Mn, Li, O, C, and~H atoms, respectively.
}
\end{figure}

AIMD potential-of-mean-force (PMF, denoted as $W(R)$ hereafter) calculations
apply composite reaction coordinates of the form $R$=$R_1$$-$$R_2$.  $R$ covers
the first two steps of the reaction.  $R_1$ and $R_2$ are the distances
between the O$_{\rm E}$ and C$_{\rm C}$ atoms and between C$_{\rm C}$
and O$_1$ (on the oxide surface), respectively (Fig.~\ref{fig1}).  In the
next step, we apply another composite coordinate $R'=R_1'-R_2'$, where
$R_1'$ and $R_2'$ are the distances between H and C$_{\rm E}$, and between
H and O$_2$, respectively (Fig.~\ref{fig1}).  Harmonic potentials of
the form $B_o(R-R_o)^2$ and a series of 10 windows with a progression of
$R_o$ values span the reaction path $R$.  $B_o$ ranges from 2~to 4~eV/\AA$^2$. 
Less windows are used along the $R'$ coordinate.  The statistical
uncertainties in $W(R)$ and $W(R')$ are estimated by splitting the trajectory
in each window into four, calculating the standard deviation, and
propagating the noise across windows assuming gaussian statistics. 

To determine Mn charge states, we examine the electron spin on each
Mn ion demarcated according to PAW orbital projections.\cite{vasp1}
All calculations allow unconstrained spin polarization.
Bulk LiMn$_2$O$_4$ crystal is antiferromagnetic (AFM).  We have imposed
an AFM ordering on alternate Mn planes along the (011) direction.  
The bare spinel surface slab is also AFM.

\section{Results}
\label{results}

\subsection{Gas phase energetics}
\label{gas}

The DFT/PBE functional predicts that gas phase disproportionation of
ethylene carbonate into carbon dioxide and acetaldehyde,
\begin{equation}
{\rm EC} \rightarrow {\rm CO}_2 + {\rm CH}_3{\rm CHO} , \label{eq1}
\end{equation}
is exothermic by 0.317~eV at T=0~K, not counting zero-point energy or finite
temperature entropic effects which should further favor dissociation.  This
result is qualitatively consistent with a reported hybrid DFT predictions
which however
include entropy contributions.\cite{assary}  We have not calculated the
activation energy, but it must be high for metastable EC to be useful in
battery operations.  No electron is transferred in Eq.~\ref{eq1}.  The
corresponding disproportionation products for PC would be CO$_2$ and acetone
((CH$_3$)$_2$CO).  Therefore observation of CO$_2$ and acetone/acetaldehyde in
experiments may not always arise from electrolyte oxidation, as was already
pointed out.\cite{ufheil,joho}  They could be also be the product of
EC/PC catalytic decomposition on battery material surfaces.

\subsection{Reaction intermediates}

\label{intermediates}

Next we consider the intermediates associated with EC decomposition on
Li$_{\rm 0.6}$Mn$_2$O$_4$.  Discussions of the activation barriers between
them will be deferred to later sections.

Figure~\ref{fig2}a (configuration ``A'') depicts an intact EC physisorbed
on the (100) surface.   The carbonyl oxygen (O$_{\rm C}$) is the only atom
strongly coordinated to the oxide, and is 2.16~\AA\, from one of the two
surface Mn(IV) ions (among eight surface Mn, appendix A) in the simulation
cell.  The overall adsorption energy is 0.480~eV.  This tilted EC geometry
binds more strongly to the surface than one where EC is in upright.  The
overall magnetic polarization of the simulation cell is 2~$\mu_{\rm B}$,
If a net zero spin polarization constraint is imposed, the total energy
increases by only 0.040~eV.  This underscores the fact the energy
differences associated with spin flipping tend to be small.  

In the first reaction intermediate~B (Fig.~\ref{fig2}b), the carbonyl
carbon atom (C$_{\rm C}$, Fig.~\ref{fig1}) is subjected to nucleophilic
attack by an oxygen atom on the surface, becoming 4-coordinated and $sp^3$
hybridized.  This bent EC species is reminscent of the bent EC$^-$ geometry
during electrochemical {\it reduction} of EC at anode surfaces.\cite{ald}
However, a maximally localized Wannier function\cite{wannier}
analysis performed on this configuration reveals that the EC carries no net
charge.  The configuration change is slightly exothermic (Table~\ref{table1}).
Note that two types of oxygen ions exist
on the (100) surface even in the absence of adsorbed EC: those bonded to
an Mn ion on the subsurface layer, and those that are not.  The latter
are more reactive, and the EC adsorption configuration is chosen so
that the C$_{\rm C}$ atom can bind to the latter type.  

\begin{table}\centering
\begin{tabular}{||l|r|r|r|r|r||} \hline
method & A & B & C & D & E \\ \hline
DFT+U & 0.000 & -0.011 & -0.100 & -2.057 & -2.144\\
PBE0  & 0.000 & -0.107 & -0.260 & -1.911 & NA\\ \hline
\end{tabular}
\caption[]
{\label{table1} \noindent
Energies of configurations A-D (Fig.~\ref{fig2}), in eV, computed
at T=0~K using different methods.
}
\end{table}

The next intermediate~C (Fig.~\ref{fig2}c) features one broken
C$_{\rm C}$-O$_{\rm E}$ bond, with the C$_{\rm C}$ atom now reverting
to $sp^2$ hybridization.  Both the O$_{\rm E}$ atom on the broken
bond and C$_{\rm C}$ become coordinated to the surface while the carbonyl
oxygen (O$_{\rm C}$) is detached.  The surface Mn ion bound to the
O$_{\rm E}$ atom now has 6-ionic or covalent bonds and transitions from
a $+3$ to a $+4$ charge state, with an $3d$ electron originally in its $3d$
shell migrating to a Mn(IV) on the surface (appendix~A).

In Fig.~\ref{fig2}d (product~D), the proton on the C$_{\rm E}$
atom closest to the surface is transferred to a surface oxygen
with a substantial gain in energy (Table~\ref{table1}).
Counting all Wannier orbitals centered closer to the EC~C and~O
nuclei than Li$_{0.6}$Mn$_2$O$_4$ atoms as belonging to the EC fragment,
EC now contains 32 valence $e^-$, implying that two of its $e^-$ have been
transferred to the oxide (appendix~A).  The oxidized EC fragment contains a
$+33|e|$ pseudovalent nuclear charge and a $+|e|$~net charge.  This
oxidative step is apparently facilitated by the preceeding
bond making/breaking events.

The proton-accepting O~ion on the oxide surface is coordinated to 3 other
Mn ions.  The average Mn-O distance associated with this O atom becomes
2.13~\AA, an increase
of 0.14~\AA\, from the value (1.99~\AA) prior to proton transfer.
Thus EC oxidation on the spinel oxide surface has led to significant
weakening of the Mn-O ionic bond network on the Li$_x$Mn$_2$O$_4$ surface.
We argue that this may contribute to Mn(II) dissolution\cite{pasquier}
(see also appendix B).

Finally, as will be discussed in Sec.~\ref{c_to_d},
the configuration in Fig.~\ref{fig2}d spontaneously reorganizes
during AIMD simulations of the liquid EC/Li$_{0.6}$Mn$_2$O$_4$ interface.
An oxygen ion on the oxide surface is pulled outwards, and the O$_{\rm C}$
atom re-coordinates to an adjacent surface Mn ion.  Extracting the EC
fragment configuration from an AIMD snapshot, transplanting it into the
smaller UHV simulation cell to give a cell commensurate
with those of Figs.~\ref{fig2}a-d, and optimizing the geometry
yield Fig.~\ref{fig2}e, with a further 0.087~eV gain in energy.
Thus all steps of the reaction are exothermic or approximately thermoneutral
(Table~\ref{table1}).  The configuration in Fig.~\ref{fig2}e maintains a
$+|e|$ charge in both the T=0~K optimized geometry and the AIMD snapshot,
counting only the electrons on the original C and O atoms of the EC.

DFT+U predictions depend on the value of $U$.
This is especially a concern with the C$\rightarrow$D reaction,
a redox process accompanied by Mn charge state changes.  The redox
potentials of transition metal ion centers have been shown to depend on
DFT functionals in some cases.\cite{cop_paper,redox1,redox2,redox3}
Thus we have applied the widely used hybrid DFT functional PBE0, which
carries no tunable parameters, to spot-check DFT+U results.  The
predicted exothermicities differ by less than 0.16~eV in all cases considered
(Table~\ref{table1}).  In particular, the overall A$\rightarrow$D redox
reaction at T=0~K differ only by 0.15~eV.  The good agreement shows that
the results are not strongly dependent on DFT functionals, and gives
us confidence in DFT+U-based predictions for our system.

\subsection{Activation (free) energies: A to C}

Figure~\ref{fig3} depicts the zero temperature energy profile
connecting A (Fig.~\ref{fig2}a) and~B (Fig.~\ref{fig2}b),
computed using the climbing image NEB method.  The barrier is
0.240~eV.  This small barrier is substantially less than the
endothermicity associated with the EC bending motion on hydroxylated
LiAlO$_2$ surfaces,\cite{ald} presumably because EC binds more 
strongly to Li$_x$Mn$_2$O$_4$ (100).

The barrier between intermediates~B and~C has not been calculated
at T=0~K.  Instead, AIMD simulations of liquid EC/Li$_x$M$_2$nO$_4$
interfaces at finite temperature cover the A$\rightarrow$C steps.
But first we discuss unconstrained AIMD simulations of EC liquid
confined between Li$_x$Mn$_2$O$_4$ surfaces prior to any chemical
reactions.  Intact EC molecules have large dipole moments but are
generally much more strongly coordinated to cations than anions.
The 8~surface Mn ions are observed to be coordinated to
between 2~and~7 EC molecules, where we have used the arbitary but
reasonable coordination criterion that Mn must be within 2.5~\AA\,
of one of the EC oxygen atoms.  The mean value is 3.7 over the trajectory.
As will be shown, EC should decompose on a subsecond timescale on
Li$_{0.6}$Mn$_2$O$_4$ (100).  Therefore the distribution of intact liquid
EC on the clean (100) surface is of academic interest.\cite{grimme}
The primary purpose of including the EC liquid is to provide a reasonable
EC liquid solvation environment.

PMF calculations are initiated from the unconstrained trajectory.  We start
with a snapshot where an intact surface EC molecule exhibits an adsorption
geometry resembling Fig.~\ref{fig2}a.  As mentioned in Sec.~\ref{method},
a composite reaction coordinate $R$=$R_1$-$R_2$ is applied, where $R_1$
is the distance between C$_{\rm C}$ and O$_{\rm E}$ and $R_2$ is that
between $C_{\rm C}$ and O$_1$ on the surface (Fig.~\ref{fig1}).  If the simple 
coordinate $R=R_1$ is used, the EC tends to desorb from the surface during
the AIMD trajectory before bond breaking can occur.  If $R=R_2$, the
reaction stops after configuration~B is reached.

Figure~\ref{fig4}a depicts the $W(R)$ computed using this composite
coordinate.  The minimum associated with intermediate~B is less stable
than the reactant by 0.24$\pm$0.05~eV.  This is less favorable than the
$-0.011$~eV found at T=0~K (Table~\ref{table1}).  The barrier between
A~and~B also increases to a still modest 0.48$\pm$0.05~eV.  The reason is most
likely the entropy change which is included in $W(R)$ but not in
Fig.~\ref{fig3}a.  The 4-coordinated intermedate~B is evidently more
constrained and entropically unfavorable than the loosely bound, flat EC
molecule (Fig.~\ref{fig2}a).  

The AIMD-predicted barrier between intermediates~B and~C is smaller
than that between A~and~B.  The exothermicity between the $W(R)$ minima
associated with C~and~A is $-0.06$$\pm$0.06~eV, which is similar to the
energy difference computed at T=0~K ($-0.10$~eV).  Thus, the main
conclusion is that the liquid EC environment does not qualitatively modify
the conclusions obtained under UHV conditions.

\subsection{Activation (free) energies: C to E}
\label{c_to_d}

Figure~\ref{fig4}a depicts the T=0~K energy profile between
intermediates~C and~D.  The reaction is exothermic by 1.957~eV
and exhibits a 0.618~eV energy barrier.  Ignoring entropic
effects and assuming a typical molecular vibrational prefactor
of 10$^{12}$/s, such an enthalpic barrier permits a 18/s reaction
rate.  As a typical battery charging/discharging rate is 1~C
(i.e., takes about an hour), the predicted subsecond reaction rate
means that EC oxidation proceeds readily during battery operations.

Next we consider temperature/solvent effects.  Figure~\ref{fig4}b depicts the
AIMD $W(R')$ for this segment of the reaction.  We apply a second composite
reaction coordinate $R'=R_1'-R_2'$, where $R_1'$ and $R_2'$ are the distances
between H and C$_{\rm E}$, and between H and O$_2$, respectively
(Fig.~\ref{fig1}a; Sec.~\ref{method}).  The predicted liquid state $W(R')$
barrier (0.54~eV) is slightly smaller than the gas phase value
(Fig.~\ref{fig4}a).  The overall reaction exothermicity may be higher
in the presence of liquid EC than under UHV conditions because the product
is charged and stabilized by the high dielectric liquid environment.
As the barrier is the main object of interest, we have not extended
$W(R')$ far into the product channel.  It is sufficient to know that
the reaction is strongly exothermic.

Note that the electron transfer from C~to~D occurs right at the
electrode-solvent interface.  It is distinct from the long-range
electron transfer mechanism experienced by ``outershell'' ionic
complexes in classic electrochemical studies, and should consequently
be much less influenced by the electric double layer.

Configuration~D spontaneously reorganizes to~E (Fig.~\ref{fig2}e) after
a 2~ps AIMD trajectory.  We have not computed the free energy barrier
between these configurations, but it must be on the order of $k_{\rm B}T$
for the conformation change to occur on such a short timescale.

\subsection{Other products/intermediates are unfavorable}

\begin{figure}
\centerline{\hbox{ (a) \epsfxsize=2.25in \epsfbox{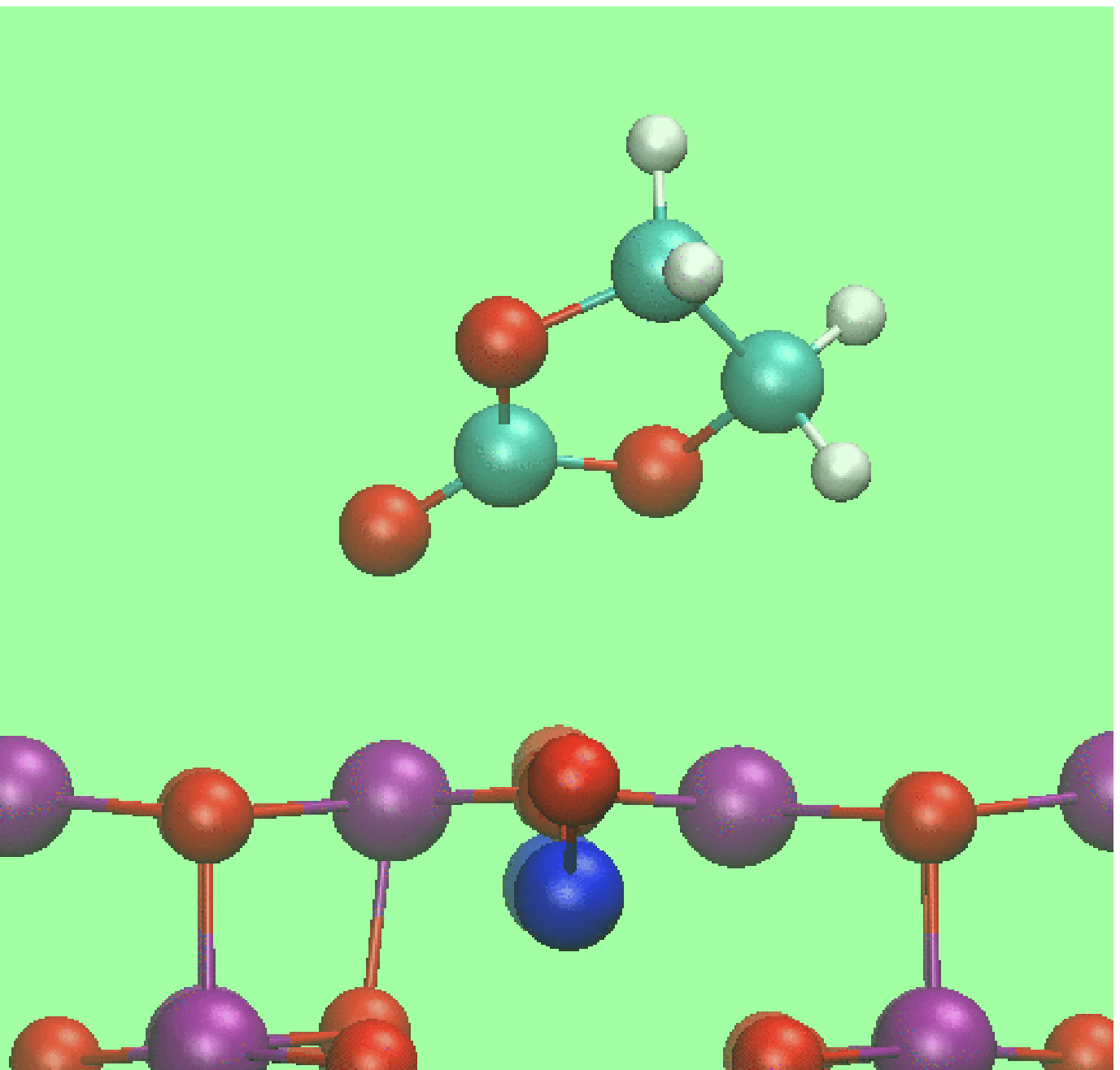}
                   (b) \epsfxsize=2.25in \epsfbox{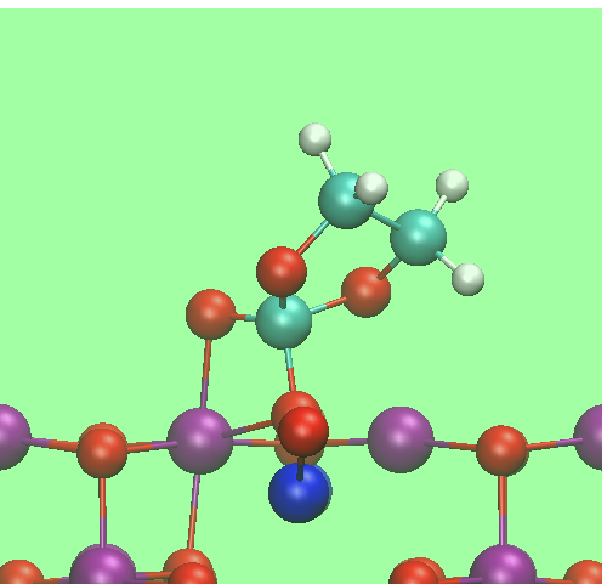} }}
\centerline{\hbox{ (c) \epsfxsize=2.25in \epsfbox{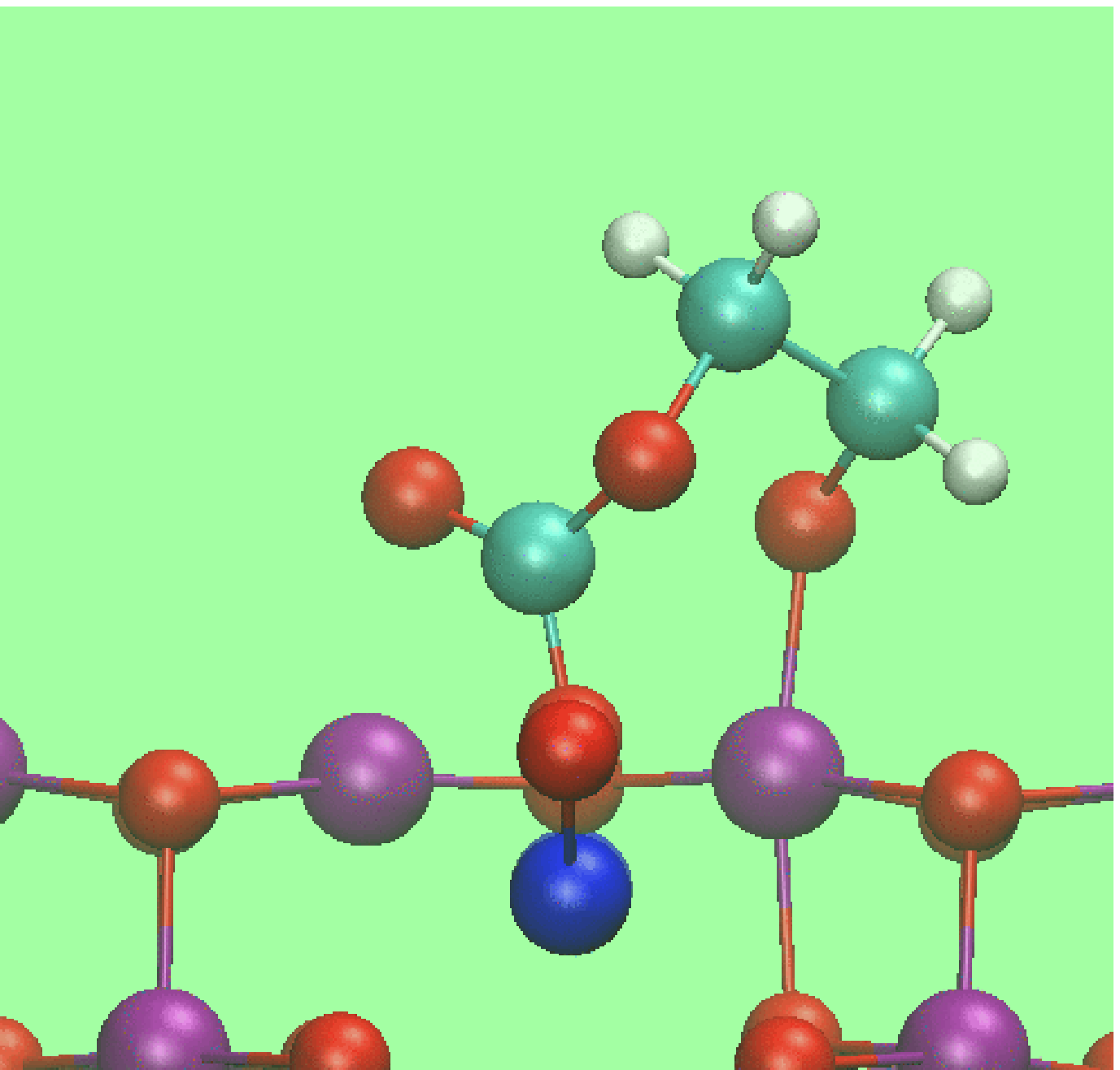}
                   (d) \epsfxsize=2.25in \epsfbox{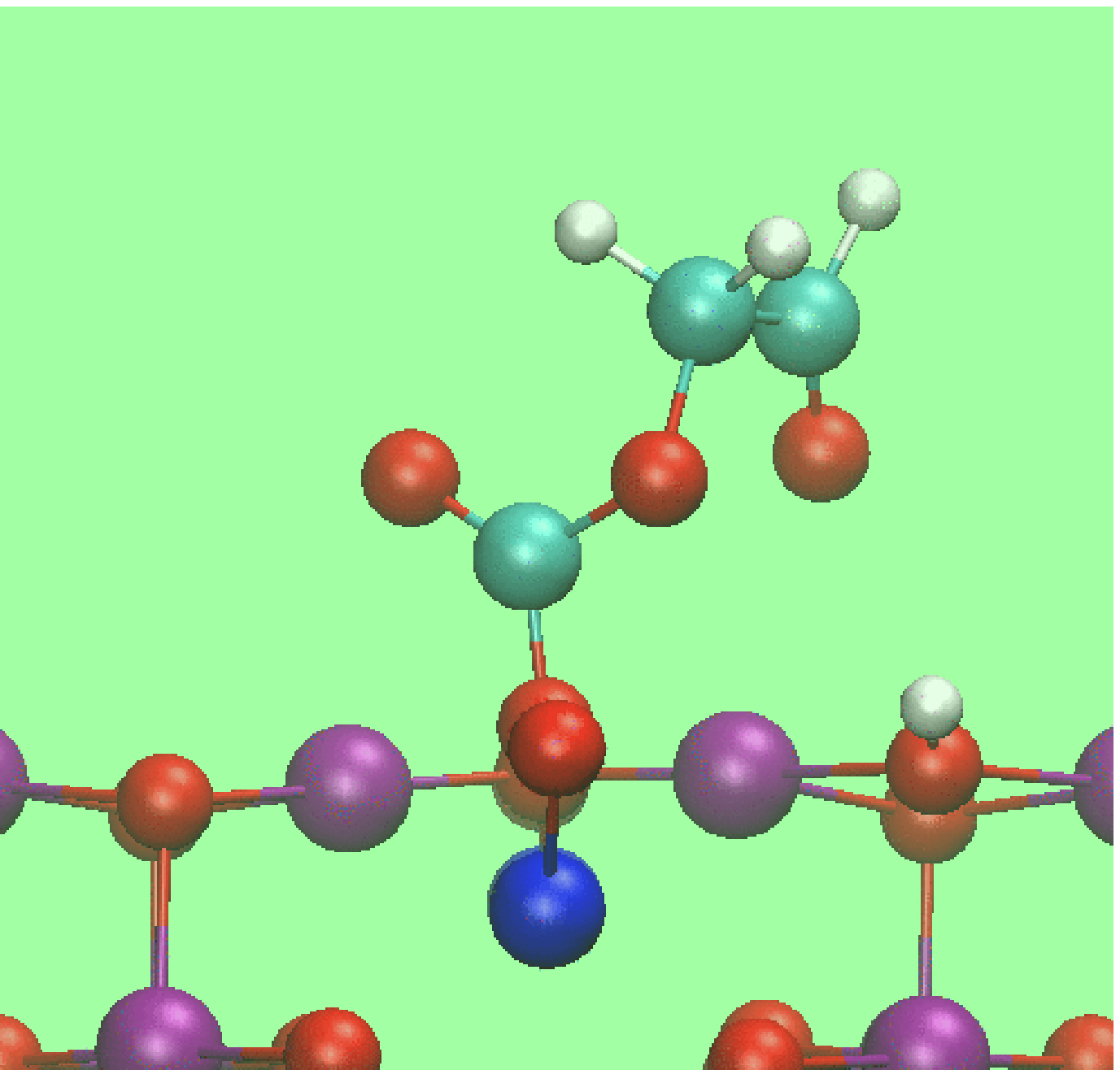} }}
\centerline{\hbox{ (e) \epsfxsize=2.25in \epsfbox{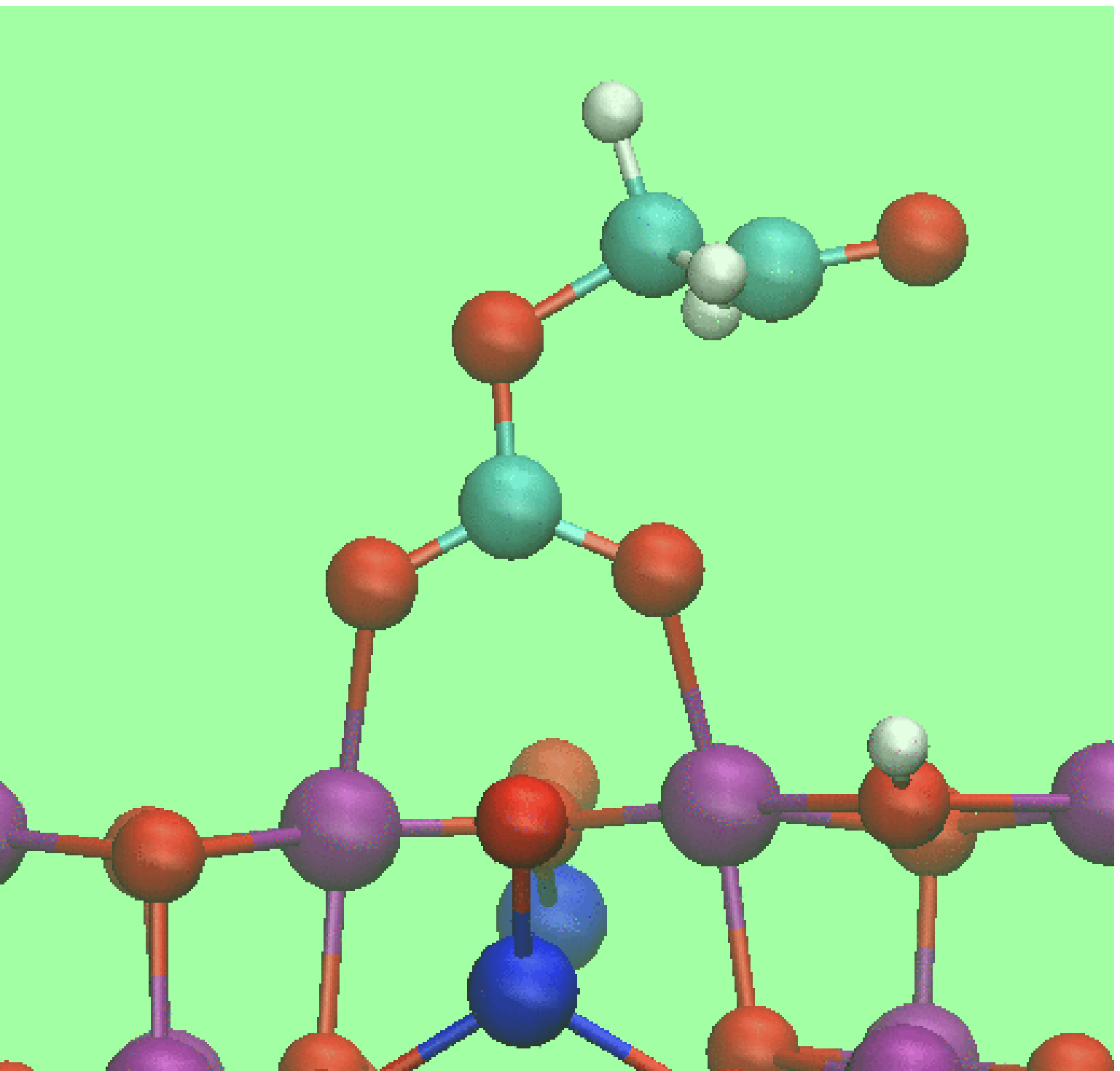}
                   (f) \epsfxsize=2.25in \epsfbox{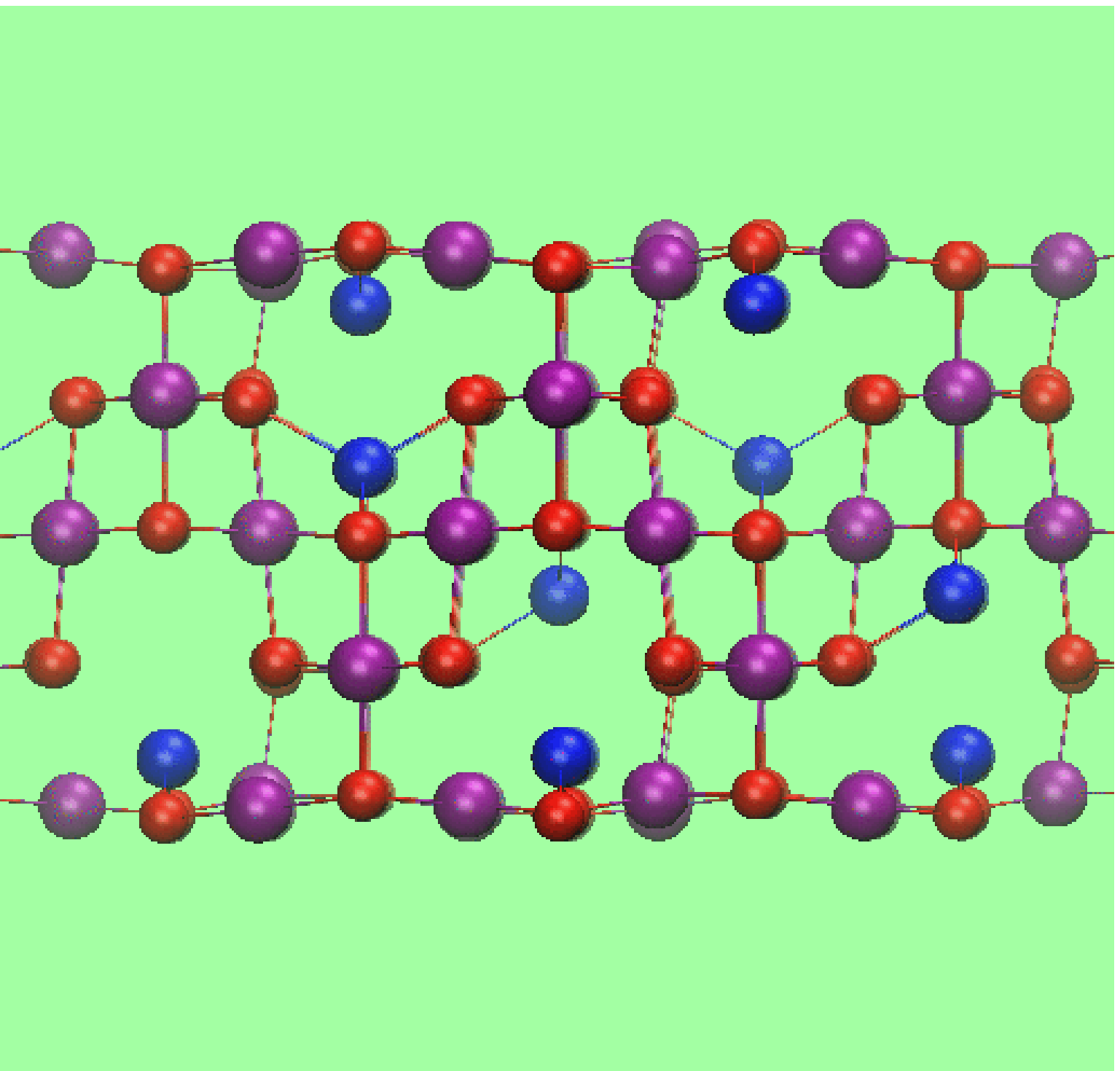} }}
\caption[]
{\label{fig2} \noindent
(a) Intact EC (A) on Li$_{0.6}$Mn$_2$O$_4$
(100) surface.  (b) Intermediate~B.  Note that the EC C$_{\rm C}$ atom
sits atop a surface oxygen ion that is not bonded to a Mn immediately
below.  (c) Intermedate~C, with a broken C$_{\rm C}$-O$_{\rm E}$ bond.
The surface Mn(III) ion coordinated to the O$_{\rm E}$ now becomes
a Mn(IV).  (d) Intermediate~D.  A proton is transfer to a surface
O~ion.  Two electrons are transferred; the surface Mn(IV) described in
panel (c) reverts to a Mn(III).  Configuration (d) reorganizes to (e) 
during AIMD simulations without further $e^-$ transfer.
(f) Most stable Li distribution in Li$_6$Mn$_{20}$O$_{40}$
(100) surface slab, viewed along the (011) direction,
about 3~unit cells abreast.  
}
\end{figure}

\begin{figure}
%\centerline{\hbox{ \epsfxsize=3.00in \epsfbox{fig3.ps} }}
\centerline{\hbox{ \epsfxsize=4.50in \epsfbox{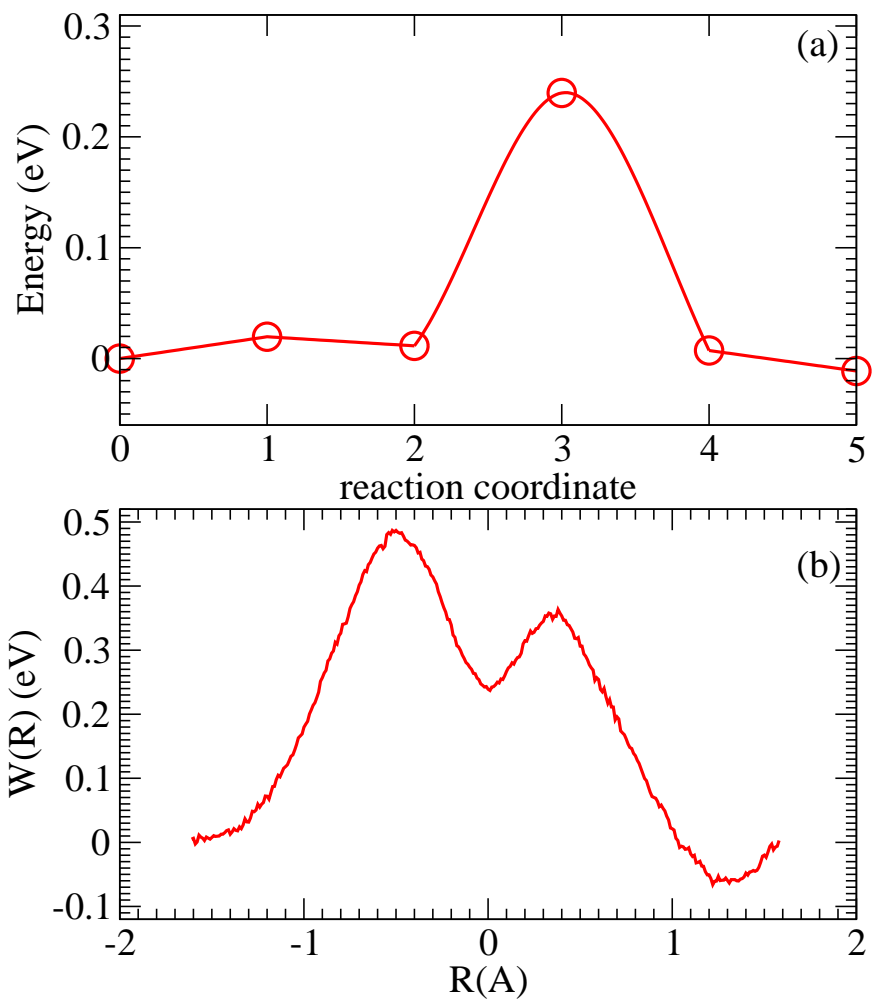} }}
\caption[]
{\label{fig3} \noindent
(a) Energy profile linking configurations A~and~B, computed at T=0~K
using the NEB method. (b) Potential-of-mean-force linking the A, B, and~C
free energy basins along the composite reaction coordinate $R$ described
in the text.
}
\end{figure}

\begin{figure}
%\centerline{\hbox{ \epsfxsize=3.00in \epsfbox{fig4.ps} }}
\centerline{\hbox{ \epsfxsize=4.50in \epsfbox{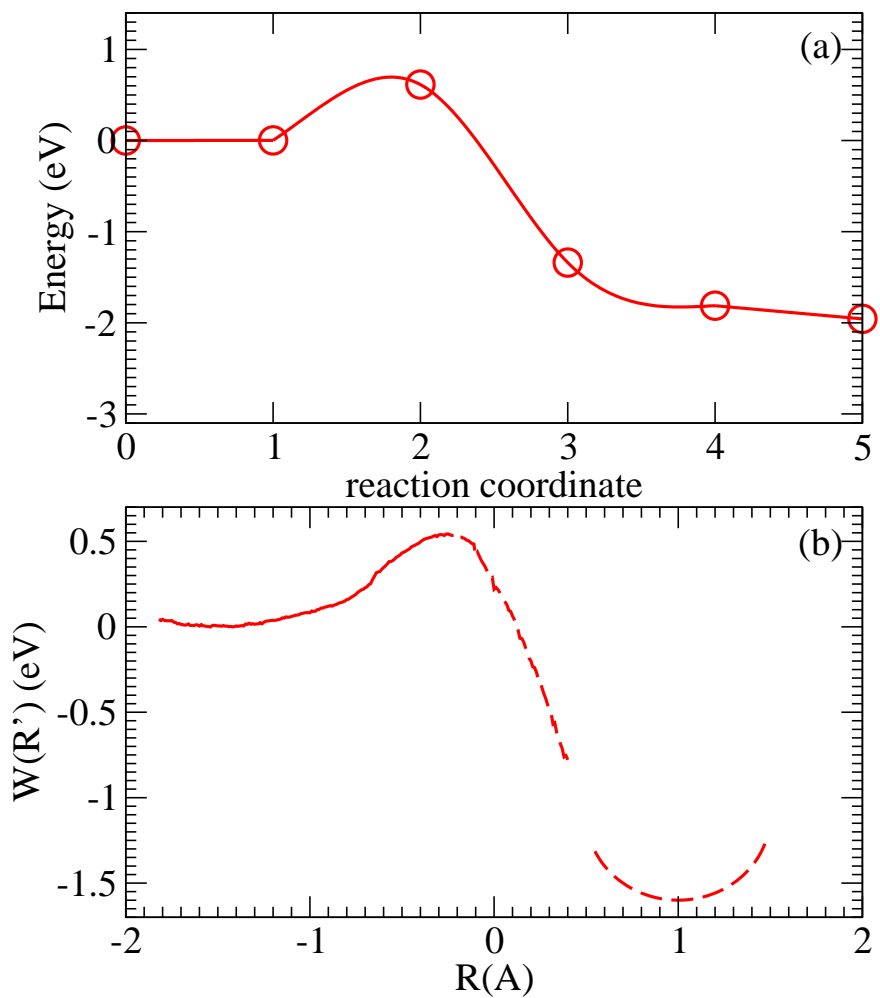} }}
\caption[]
{\label{fig4} \noindent
(a) Energy profile linking configurations C~and~D, computed at T=0~K
using the NEB method. (b) Potential-of-mean-force linking C~and~D
along the composite reaction coordinate $R'$ described in the text.
In panel (b), the bottom of the well at $R$=1~\AA\, has not been determined.
}
\end{figure}

\begin{figure}
%\centerline{\hbox{ (a) \epsfxsize=1.50in \epsfbox{fig5a.ps}
%                   (b) \epsfxsize=1.50in \epsfbox{fig5b.ps} }}
%\centerline{\hbox{ (c) \epsfxsize=1.50in \epsfbox{fig5c.ps}
%                   (d) \epsfxsize=1.50in \epsfbox{fig5d.ps} }}
\centerline{\hbox{ (a) \epsfxsize=2.25in \epsfbox{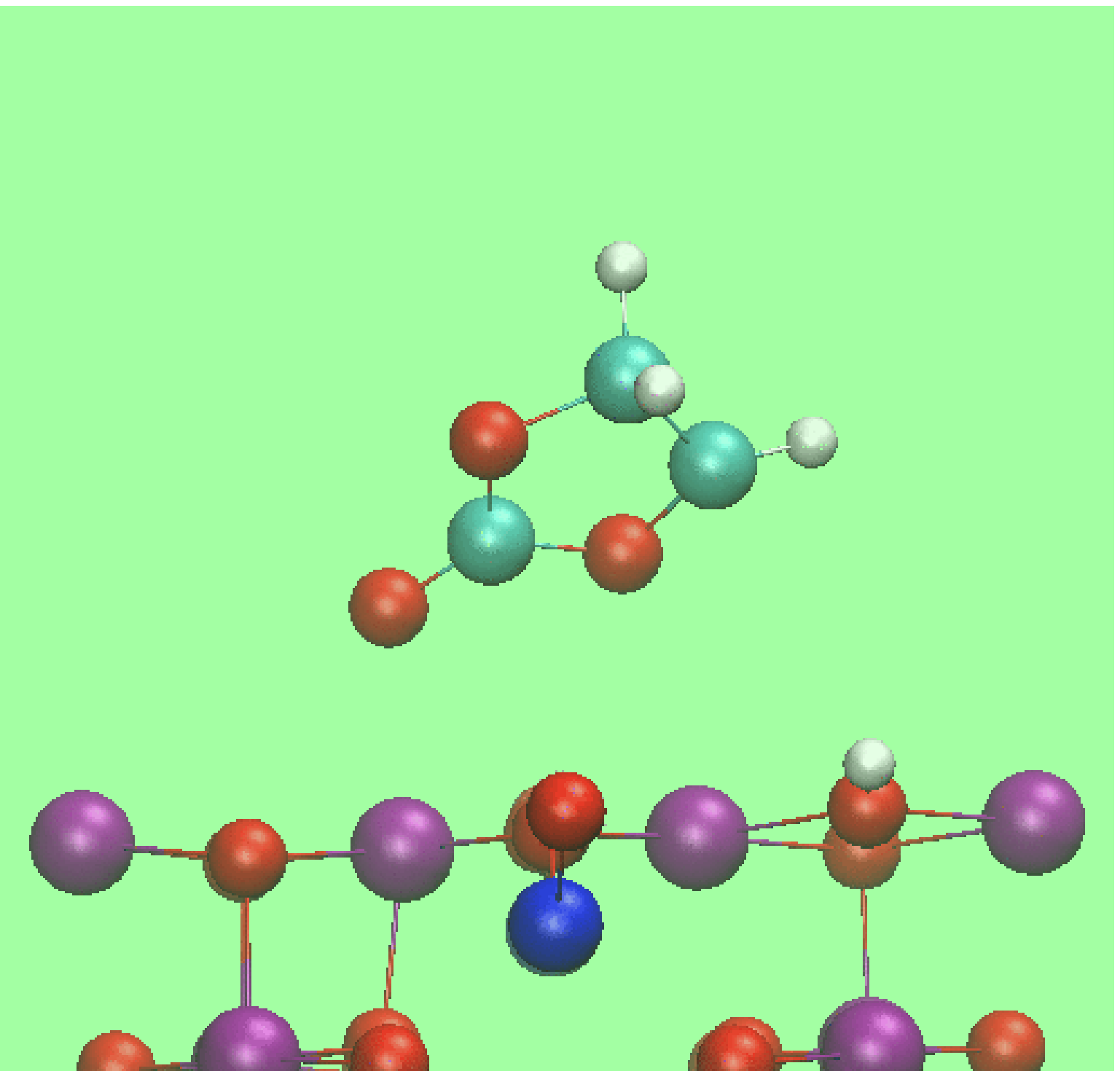}
                   (b) \epsfxsize=2.25in \epsfbox{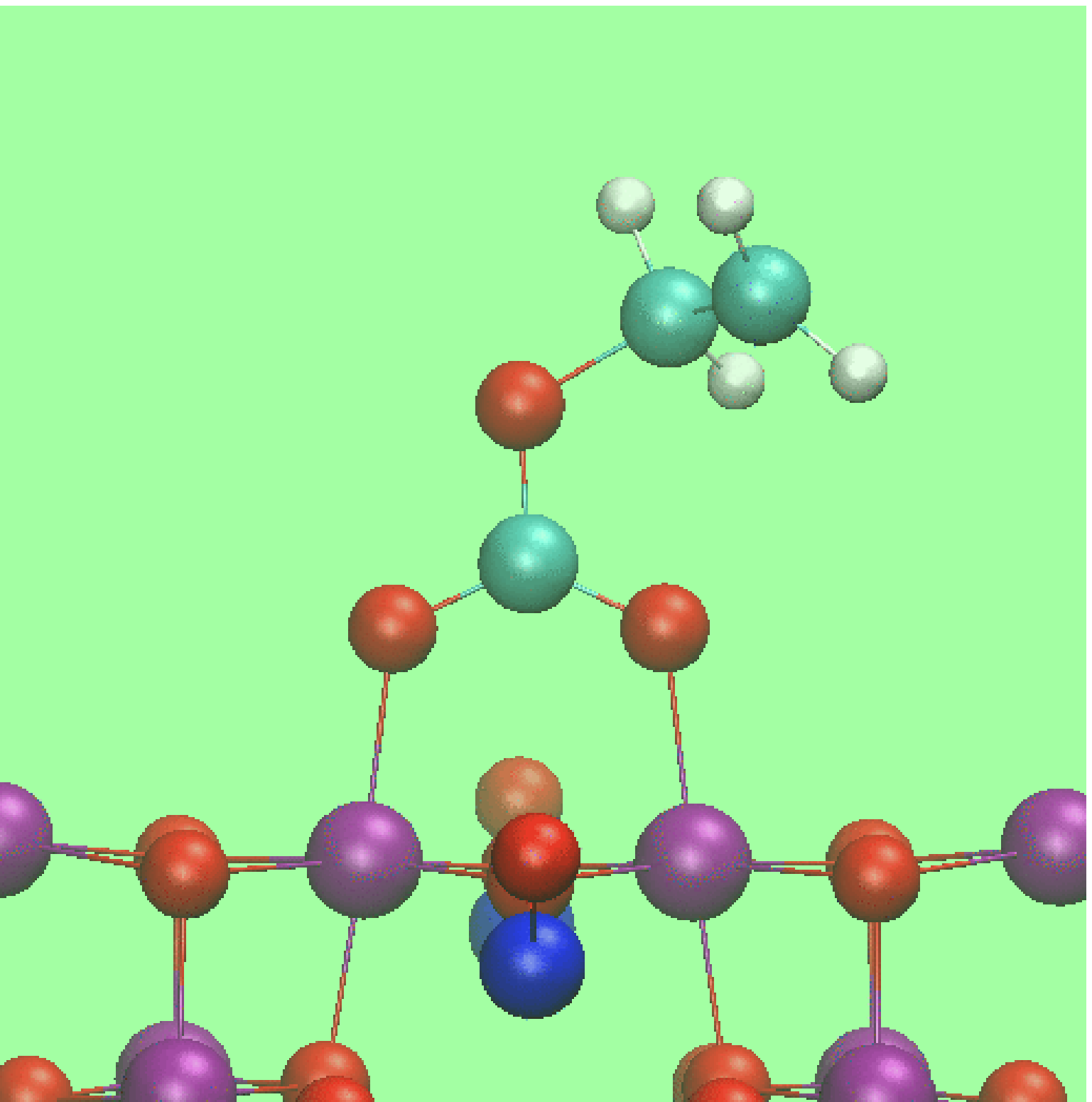} }}
\centerline{\hbox{ (c) \epsfxsize=2.25in \epsfbox{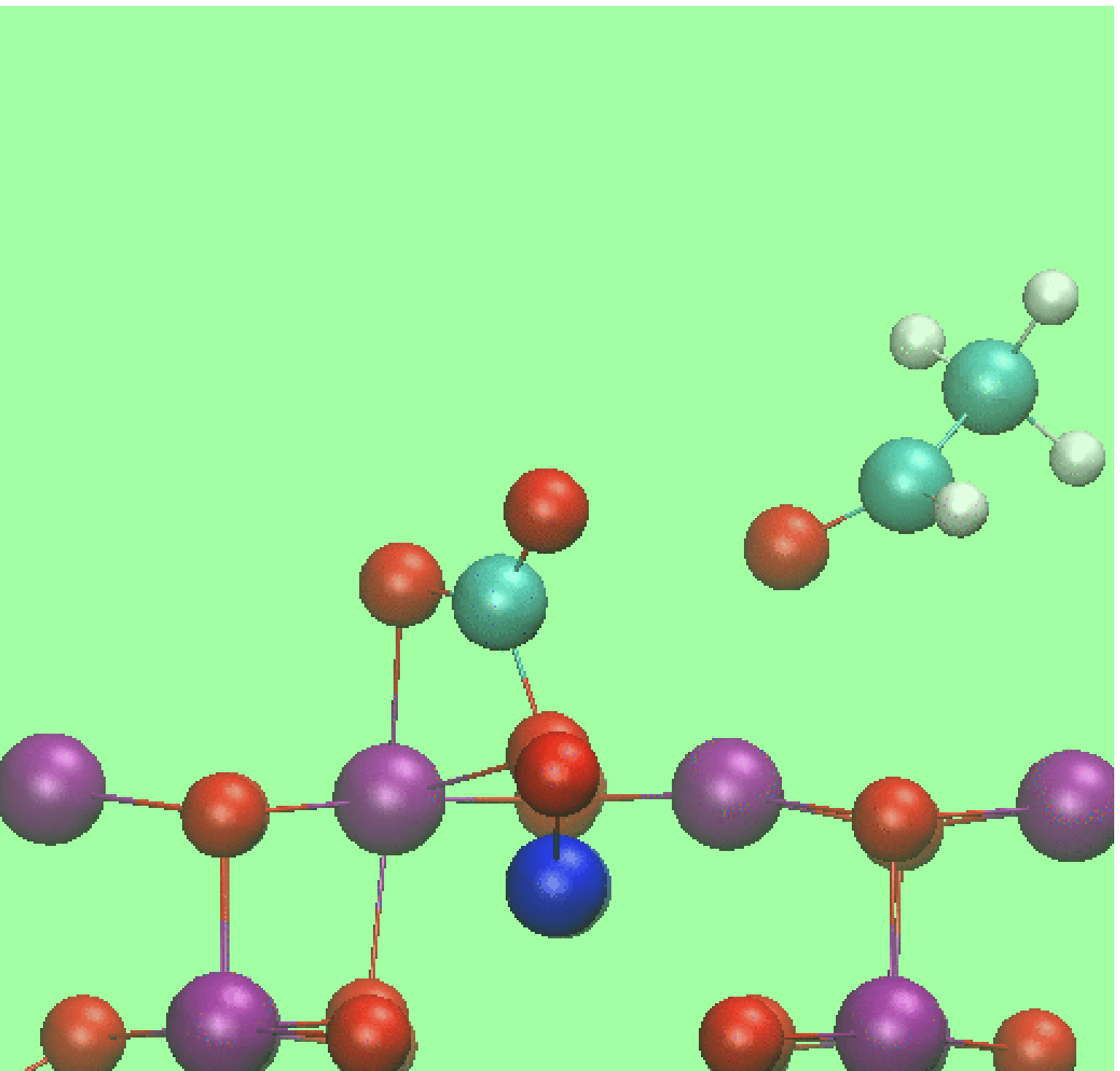}
                   (d) \epsfxsize=2.25in \epsfbox{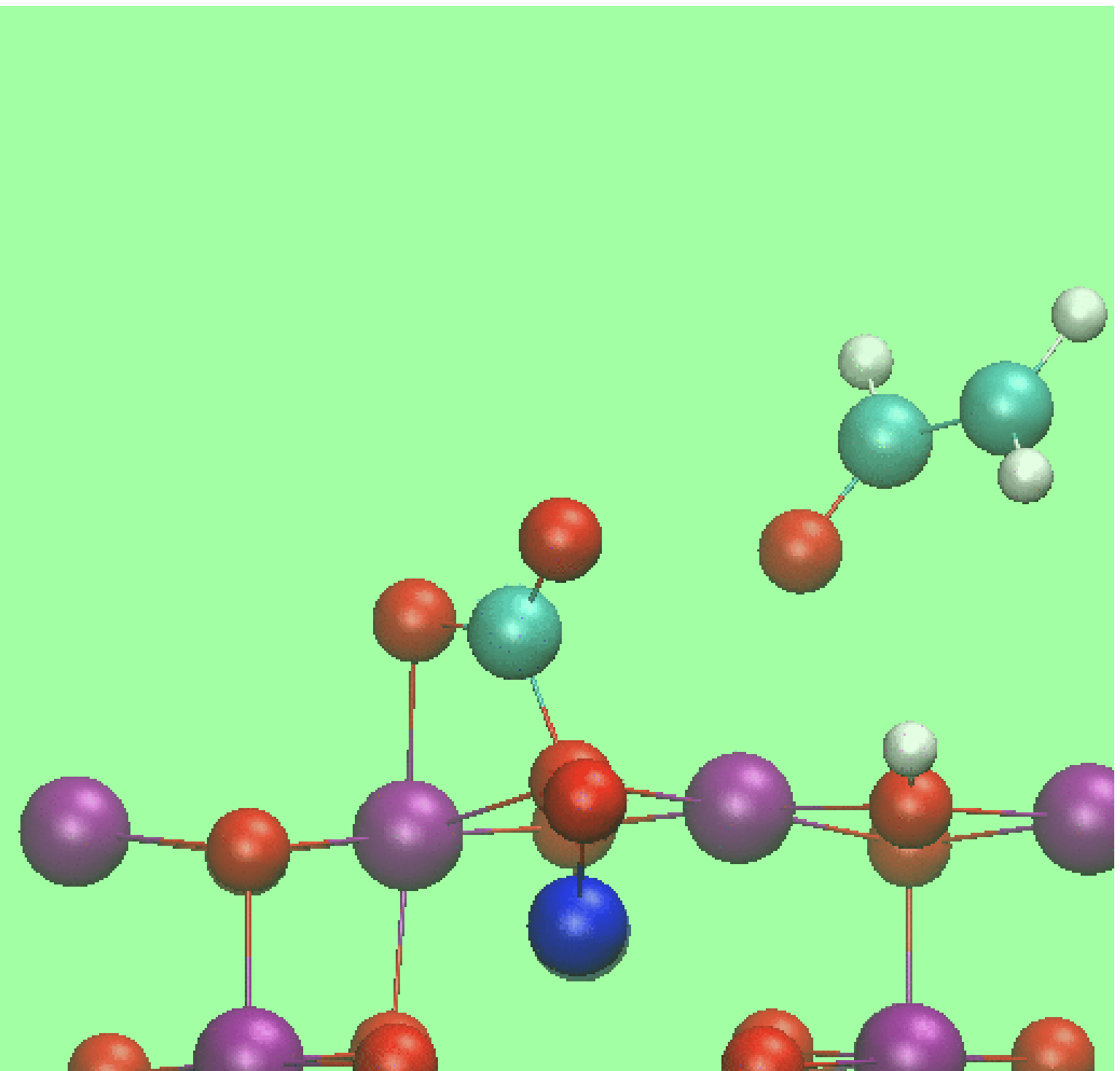} }}
\caption[]
{\label{fig5} \noindent
Four decomposition products or intermediates not considered part of
the main EC decomposition pathway,
with zero temperature energies of $+0.398$~eV, $+1.562$~ev, $-0.646$~eV,
and $-0.376$~eV relative to Fig.~\ref{fig2}a, respectively.
}
\end{figure}

Fig.~\ref{fig5}a depicts an otherwise intact EC molecule with a proton
transferred to the oxide surface.  Compared to the intact adsorbed EC
(Fig.~\ref{fig2}a), this configuration is endothermic by 0.398~eV.  In
contrast, removing a proton from the EC with a broken C$_{\rm C}$-O$_{\rm E}$
bond (Fig.~\ref{fig2}c) is exothermic by 2.057~eV overall.  This shows that
the non-oxidative, bond-breaking first steps strongly facilitate redox
reactions on the spinel surface.

Several possible decomposition products, devised from breaking a
C$_{\rm E}$-O$_{\rm E}$ bond at various stages of the reaction,
lead to higher total energies and/or high barriers and are ruled out.
In Fig.~\ref{fig5}b, we attempt to break a C$_{\rm E}$-O$_{\rm E}$ bond
in the Fig.~\ref{fig2}a configuration to form a carbonate-like group
in the first reaction step.  The optimized metastable geometry is unfavorable
by 1.562~eV compared to the physisorbed, intact EC (Fig.~\ref{fig2}a).

Fig.~\ref{fig5}c depicts a configuration resulting from cleaving a
C$_{\rm E}$-O$_{\rm E}$ bond in Fig.~\ref{fig2}c without first transferring
a proton to the surface.  One product fragment, a acetaldehyde
molecule which arises from internal proton transfer among the carbon atoms
in the CH$_2$CH$_2$O fragment, is detached from the surface, while the
other, a CO$_2$ molecule, remains  strongly coordinated to the metal
oxide.  No electron is transferred.  This configuration is exothermic
by 0.646~eV compared to the intact physisorbed EC, but is
much less stable than that in Fig.~\ref{fig2}d.  (See also Sec.~\ref{gas}.)
We have not found a converged barrier smaller than 1.5~eV for this reaction.
Hence the Fig.~\ref{fig2}d product channel is much more favorable.

Figure~\ref{fig5}d explores the possibility of C$_{\rm E}$-O$_{\rm E}$
bond-breaking following proton transfer (Fig.~\ref{fig2}d).  The
resulting product is far less stable than if the C$_{\rm E}$-O$_{\rm E}$
bond remains intact (Fig.~\ref{fig2}d).

In conclusion, we have not found a favorable reaction pathway that yields
CO$_2$ and/or acetaldehyde-like fragments.  This suggests that the partially
decomposed EC fragments chemisorbed on the (100) surface (Fig.~\ref{fig2}d
\&~\ref{fig2}e) is stable under UHV conditions.  Preliminary AIMD PMF
simulations also show that breaking a EC bond in configuration~D
(Fig.~\ref{fig2}d) or configuration~E (Fig.~\ref{fig2}e) to Fig.~\ref{fig5}d
exhibits large barriers.  

However, the chemisorbed EC fragment (Fig.~\ref{fig2}e) can
potentially attack and react with other intact solvent molecules,
potentially abstracting protons and/or electrons (``H$^-$'') from them.
This has not been observed in our AIMD simulations which are
of limited duration.  In view of Refs.~\onlinecite{borodin}
and~\onlinecite{borodin1}, it is also possible that further decomposition
of the fragment can occur in the vicinity of anions in the electrolyte,
which can act as proton shuttles.

\section{Discussions}

\subsection{Relevance to battery operating conditions}

Our calculations focus on the clean Li$_x$Mn$_2$O$_4$ (100) surface, which
has been the focus of almost all theoretical studies.\cite{ouyang2,benedek_new}
Thin Li$_2$CO$_3$ films are known to exist on cathode oxide
surfaces before cycling power.\cite{aurbach1,yang2010}  Since electrolyte
decomposition products are observed on these oxides surfaces, it has been
suggested that this film dissolves upon immersion in the electrolyte,
during or prior to cycling power.\cite{aurbach,hirayama,hirayama1}

The reaction products and pathways reported on (100) are not meant to
quantitively reproduce all experimental spinel facets present.  If the
adsorbed, partially decomposed EC$^+$ fragments (Fig.~\ref{fig2}d,e) on
Li$_{0.6}$Mn$_2$O$_4$ (100) are stable, they become part of the cathode
surface film and may interfere with Li$^+$ intercalation.  These products
block the reactive sites, of surface concentration 1/17.6~\AA$^2$, but do
not form a continuous film to stop electron tunneling from the cathode to
the electrolyte.  

The fact that PF$_6^-$ is found to accelerate Mn$_2$O$_4$ degradation
is not inconsistent with our findings.  PF$_6^-$ may further react with,
decompose, and removed the chemisorbed EC$^+$ fragments on the surface 
(Fig.~\ref{fig2}e) to form volatile organic products.\cite{borodin}
This may expose reactive sites on the oxide surface, allowing other EC
molecules to react and potentially deposit more protons.   This will
be considered in the future.  The H$_2$O/PF$_6^-$ pathway discussed in
the literature may also occur simultaneously with our predicted EC
H$^+$-donation mechanism.

As no counter electrode is present in the simulation cell, our AIMD
simulations represent an open-circuit condition with a Li$_{0.6}$Mn$_2$O$_4$
stochiometry which may be more related to battery storage than charging
experiments.  For larger values of $x$, Mn(IV) has less tendency to reside
on the oxide surface, and the reactivity of EC should decrease.  This may be
consistent with the observation that higher charging voltages yield more
interfacial reactions.\cite{novak3,novak4}

\subsection{Comparison with EC$^+$/PC$^+$ decomposition predictions made in
the absence of an electrode}

To some extent, our proposed EC decomposition mechanism (Fig.~\ref{fig2}a-e)
is the reverse of that of Refs.~\onlinecite{borodin} and~\onlinecite{borodin1}.
Oxidation and proton removal from EC occur in the last, not first, step.  A C-O
bond first ruptures on a EC initially physisorbed the Li$_x$Mn$_2$O$_4$
surface, facilitating subsequent redox reactions.  

As in this work, Refs.~\onlinecite{borodin} and~\onlinecite{borodin1}
emphasize the role of proton removal from organic solvent molecules during
oxidation.  All these studies suggest that trace water may not be needed
to initiate degradation of the cathode by creating acid molecules.  The final
products reported in Ref.~\onlinecite{borodin} (which does not consider
explicit electrodes) are coordinated to anions.  In our work, anions are
not present, and similar products are at best metastable.  

We predict preferential cleavage of the C$_{\rm C}$-O$_{\rm E}$
bond, not C$_{\rm E}$-O$_{\rm E}$, in agreement with Xing {\it et al.}'s
results on the decomposition of isolated EC$^+$ and PC$^+$.\cite{xing1,xing2}
The final products predicted by Xing {\it et al.} are not observed because
EC oxidation is accompanied by proton transfer to the surface herein.  

\subsection{More discussions on computational methods}

$e^-$ transfer between the EC and Mn ions, and between Mn ions, are implicitly
assumed to exhibit lower barriers than the rate-determining step in the
bond-breaking processes.  This follows from the DFT (and DFT+U) calculations,
where $e^-$ transfer events are implicitly assumed to adiabatically
follow ionic trajectories.  Non-adiabatic effects are not expected to
dominate, because non-adiabatic polaron transport in spinel Mn$_2$O$_4$
are not predicted to exhibit high barriers.\cite{ouyang1}  The
$e^-$ transfer between configurations~C (Fig.~\ref{fig2}c) and~D
(Fig.~\ref{fig2}d) is unlikely to be rate limiting either, because
the partially decomposed EC molecule is covalently bonded to the surface,
yielding robust $e^-$ conduction pathways.  While hybrid functionals are
generally more accurate than DFT+U for the C$\rightarrow$D reaction
which involves concerted electron and proton transfer,
we have found that both methods predict similar energy changes
(Table~\ref{table1}).

Zero point energy (ZPE) are not included in the energy changes reported
in this work.  Most bond-breaking events examined involve ``heavy'' second
row atoms.  In some cases, a proton is transferred from a carbon to an oxygen
atom.  As C-H and O-H vibrational frequencies are within 20\% of each other,
ZPE should not strongly modify the predicted exothermicities.  Proton
transfer barriers would be even lower if ZPE is included.

Finally, we mention that the reaction pathway (Fig.~\ref{fig2}) is chosen
through trial and error.  A more systematic approach would be to apply
metadynamics.\cite{meta1,meta2,meta3} The transition path sampling approach
can also be combined with AIMD simulations to yield more rigorous free
energy barriers.\cite{chandler}

\section{Conclusions}
\label{conclude}

This computational work focuses on the initial stages of ethylene carbonate
(EC) decomposition on Li$_{0.6}$Mn$_2$O$_2$ (100) surfaces.  In analogy with
studies of water adsorption and dissociation on transition metal oxide
surfaces, we perform calculations on clean, ideal surface models that
can potentially be replicated in future experiments.  Both ultra-high
vacuum (UHV) and solvent-immersed electrode conditions are considered.
Static (T=0~K) DFT calculations and finite temperature {\it ab initio}
molecular dynamics (AIMD) simulations are applied under these
conditions, respectively, yielding qualitatively similar conclusions. 

The first step of EC decomposition on Li$_{0.6}$Mn$_2$O$_4$ involves
C$_C$-O$_E$ bond-cleavage but no oxidation.  This slightly exothermic
bond-breaking event enables a proton and two electrons to be transferred
to the oxide surface in the next oxidative, strongly exothermic step.  The
activation energies for both steps are about 0.62~eV under UHV conditions
and 0.54~eV at electrode-electrolyte interfaces, according
to our DFT+U calculations.  Thus oxidation of EC on the (100) surface
is predicted to take place on a subsecond timescale.  The Mn-O ionic bonding
network is weakened by the EC proton transferred to the surface
as well as the adsorbed EC fragment, suggesting that solvent decomposition
and Mn and/or O~dissolution from the oxide surface may be related.
The fact that proton can come from organic solvents has been suggested by
an experiment work using tetrahydrofuran.\cite{jang1997}
Our work suggests a surface-assisted pathway for solvent oxidation and
surface film formation at the open circuit Li$_{0.6}$Mn$_2$O$_4$
voltage.  No assumption or separate calculation of spatial inhomogeneity
effect on the local dielectric constant is made in this work.

The predicted EC fragment chemisorbed on the oxide (100) surface is likely
not the final product.  Our goal is not to reproduce the final distribution
of organic decomposition products observed in battery experiments.
In particular, our calculations do not extend to the high voltage region
which seems to be required for CO$_2$, one of the proposed final
products.\cite{novak4} Instead, this work represents the perhaps necessary
first step of explicit modeling of electrolyte breakdown on oxide surfaces,
and paves the way for theoretical and experimental investigations
on other Li$_x$Mn$_2$O$_4$ crystal facets.

\section*{Acknowledgement}
 
We thank Nancy Missert, Nelson Bell, Yue Qi, Steve Harris, Kevin Zavadil, and
John Sullivan for useful discussions.  Solid state LiMn$_2$O$_4$ calculations
are initiated using a unit cell provided by Shirley Meng.
Sandia National Laboratories is a multiprogram laboratory
managed and operated by Sandia Corporation, a wholly owned subsidiary of
Lockheed Martin Corporation, for the U.S.~Deparment of Energy's National
Nuclear Security Administration under contract DE-AC04-94AL85000.  
UHV-condition calculations were supported by Nanostructures for Electrical
Energy Storage (NEES), an Energy Frontier Research Center funded by
the U.S.~Department of Energy, Office of Science, Office of Basic Energy
Sciences under Award Number DESC0001160.  This research used resources of
the National Energy Research Scientific Computing Center, which is supported
by the Office of Science of the U.S. Department of Energy under Contract
No.~DE-AC02-05CH11231. 

\section*{Appendix A: Charge states of Mn ions}
\label{magnet}

\begin{figure}
\centerline{\hbox{ (a) \epsfxsize=2.25in \epsfbox{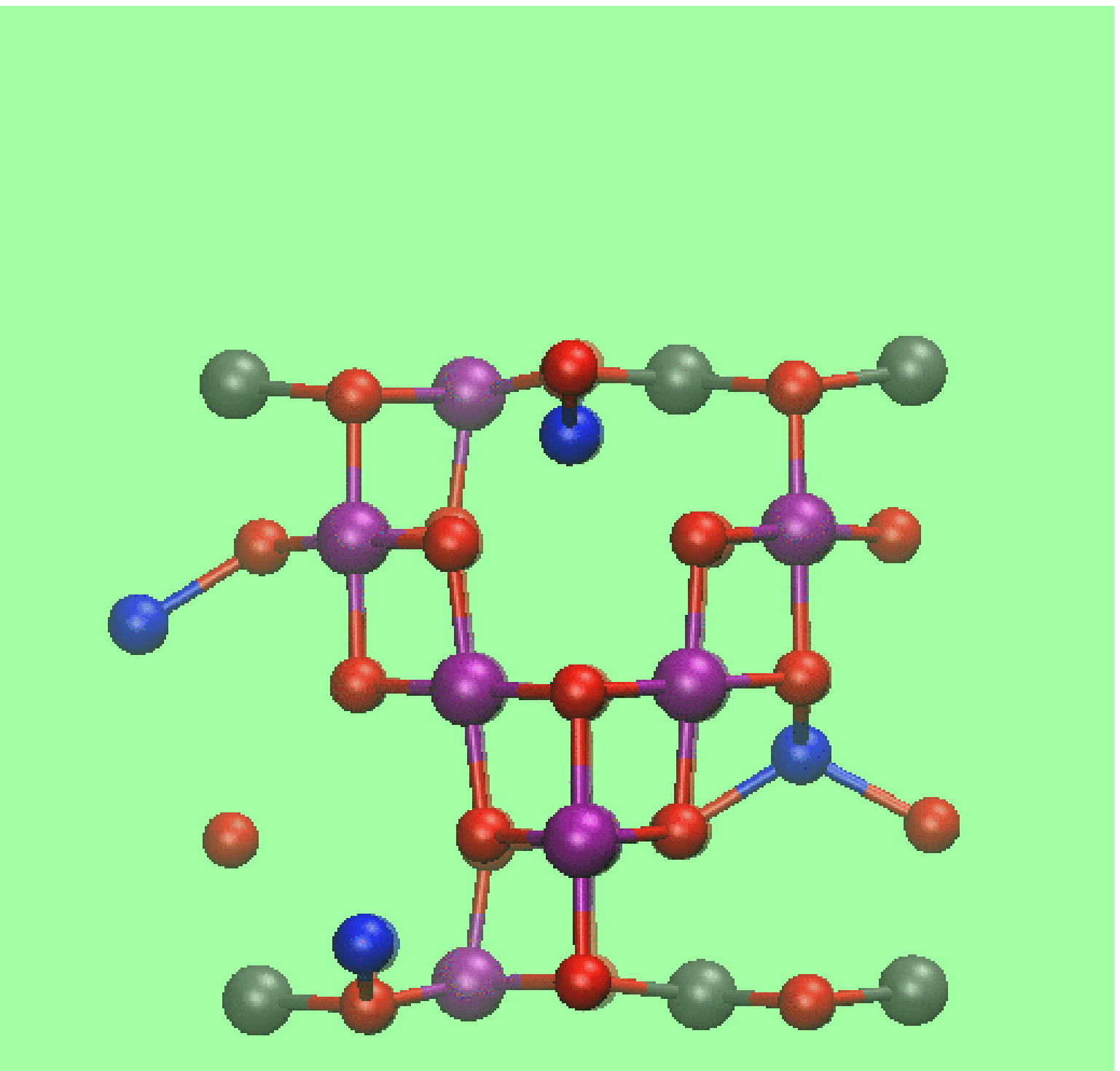}
                   (b) \epsfxsize=2.25in \epsfbox{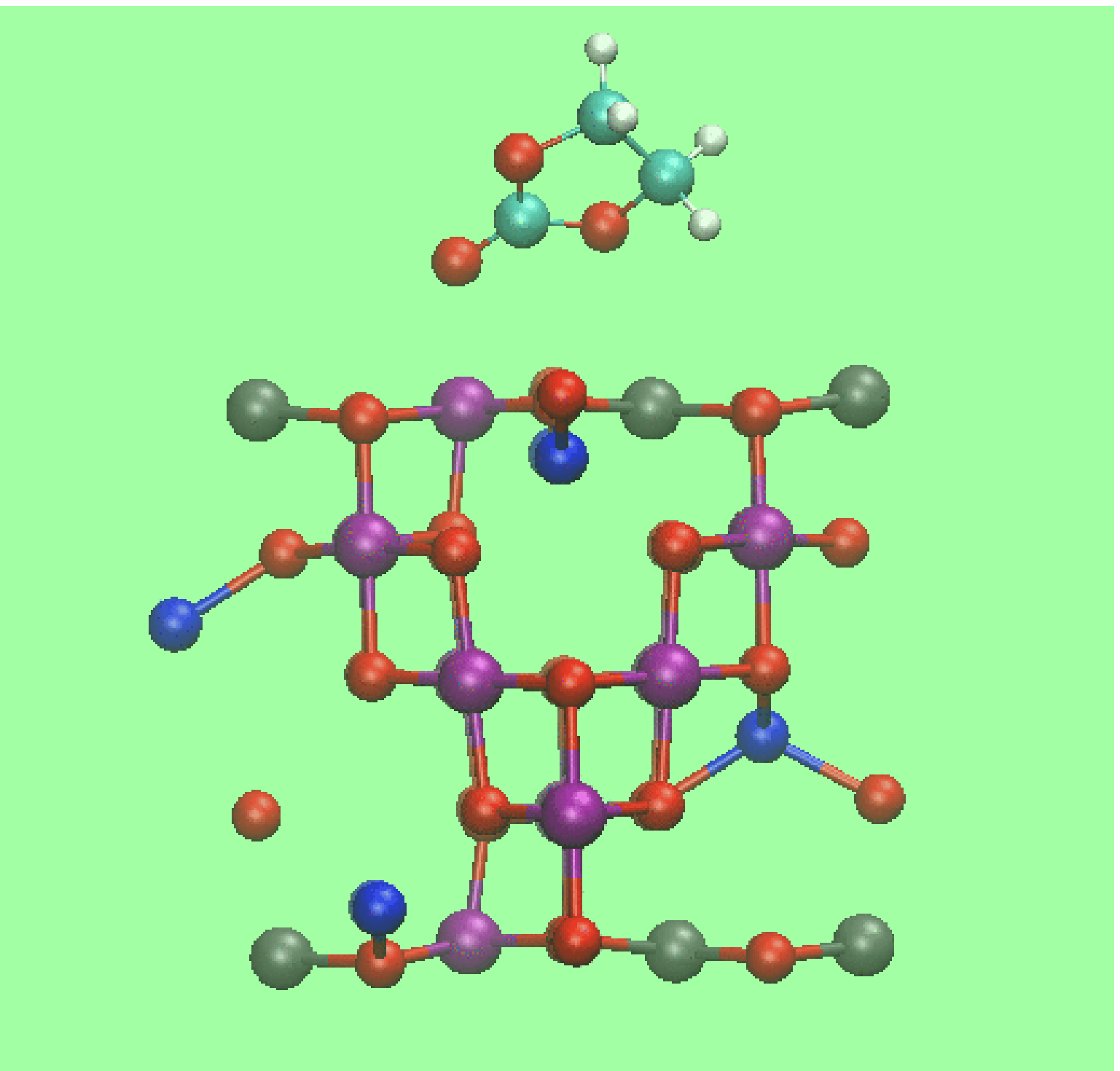} }}
\centerline{\hbox{ (c) \epsfxsize=2.25in \epsfbox{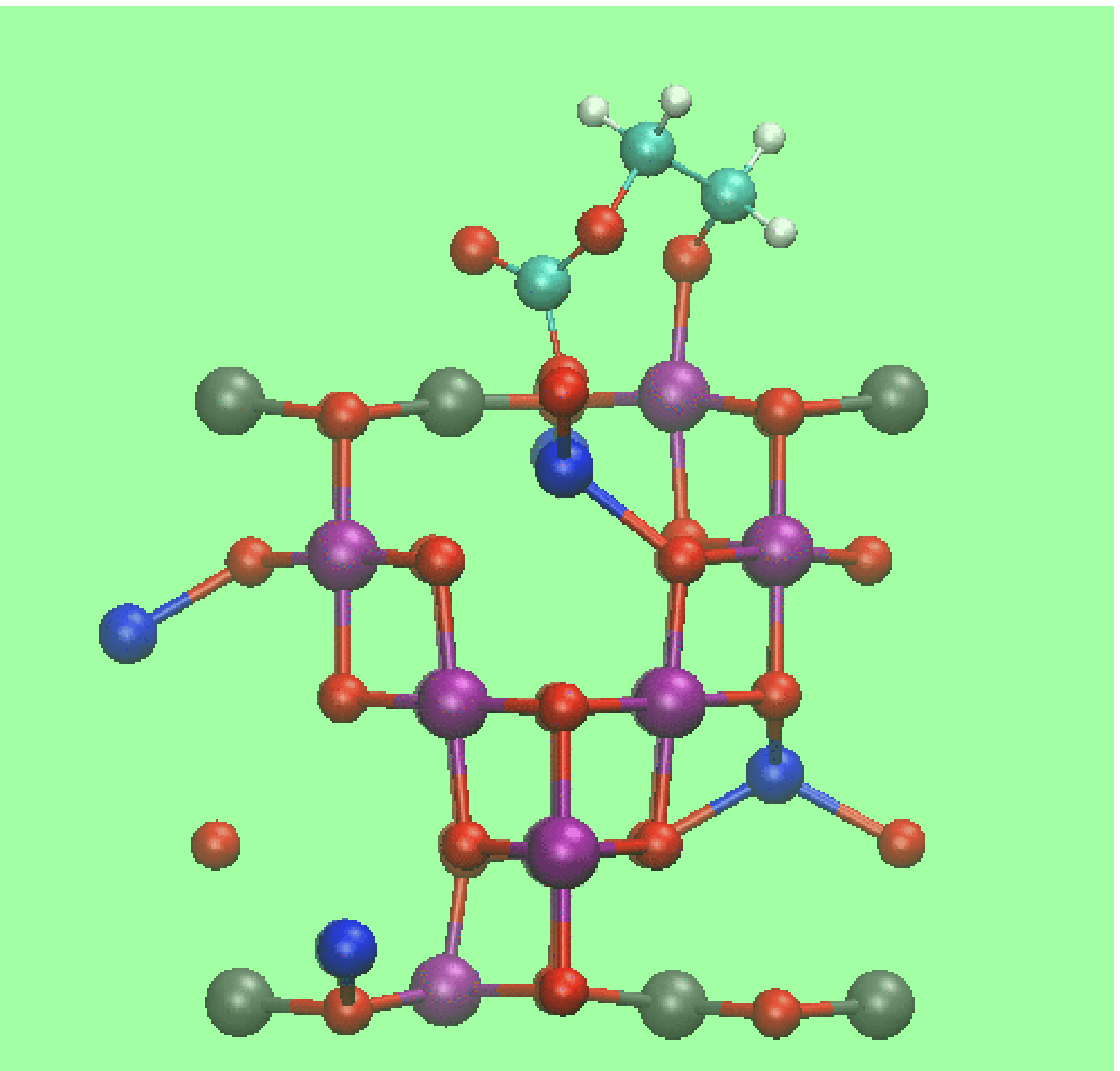}
                   (d) \epsfxsize=2.25in \epsfbox{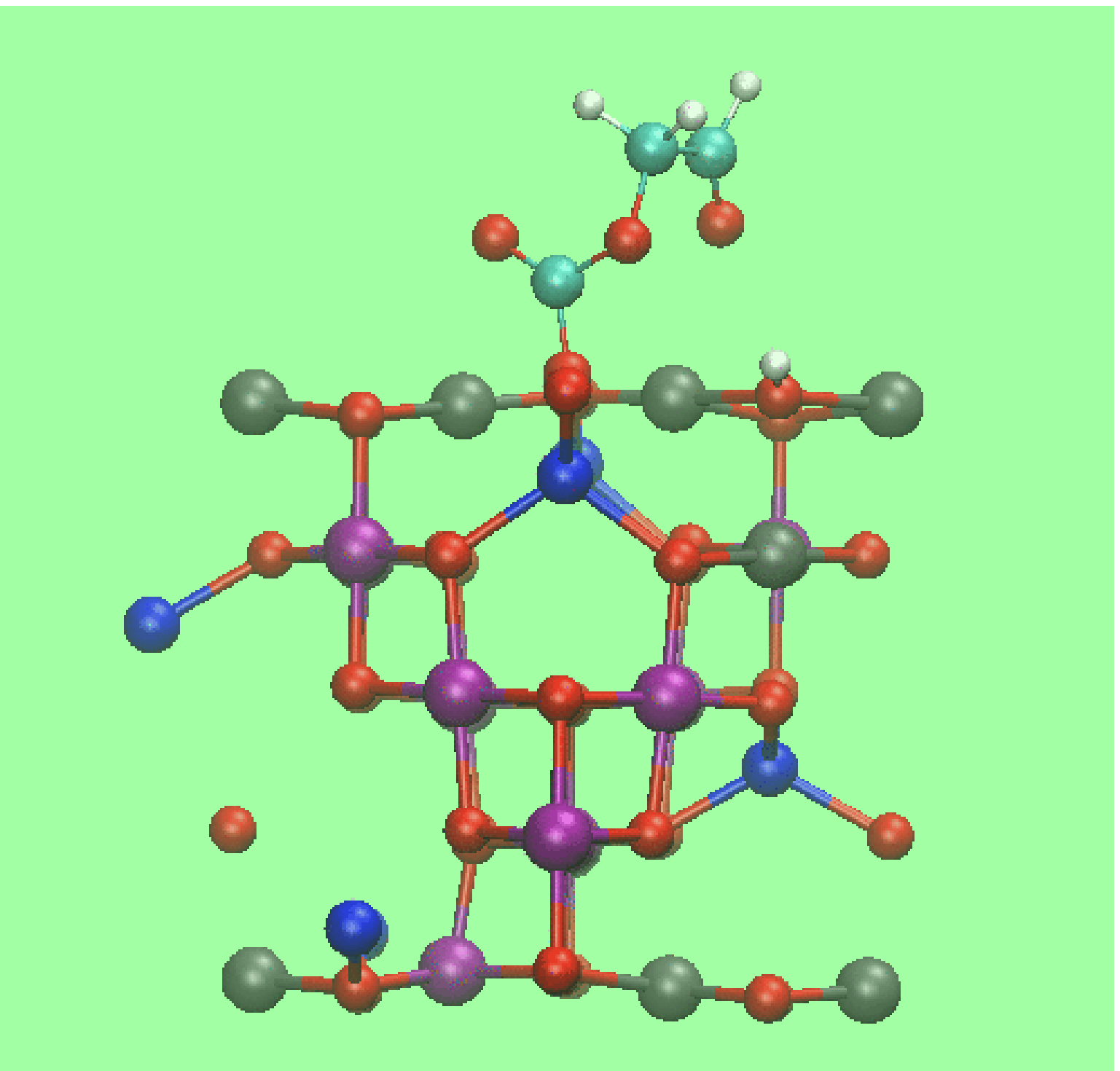} }}
\caption[]
{\label{fig7} \noindent
Magnetic moments of Mn ions corresponding to the bare Li$_{0.6}$Mn$_2$O$_4$
slab, intact EC (Fig.~\ref{fig2}a), EC with a broken C-O bond
(Fig.~\ref{fig2}c), and oxidized EC (Fig.~\ref{fig2}d).  Purple and dark-green
spheres represent Mn(IV) and Mn(III), respectively.
}
\end{figure}

In Sec.~\ref{intermediates}, changes in Mn charge states within the
oxide slab are briefly discussed.  They are explicitly depicted in this
appendix. Fig.~\ref{fig7}a shows that 6 Mn(III) and 2 Mn(IV) ions in the 
Li$_{6}$Mn$_{20}$O$_{40}$ slab reside on the (100) surface.  The
physisorption of an intact EC on a surface Mn(IV) ion does not change this
charge state distribution (Fig.~\ref{fig7}b).  Upon breaking an EC C-O bond,
however, the O~atom on that broken bond becomes strongly bound to
a surface Mn ion, which loses an electron to the adjacent, formerly Mn(IV)
ion on the surface (Fig.~\ref{fig7}c).  The oxidative step transfers
two $e^-$ to the oxide slab.  As a result, both a surface and an interior
Mn(IV) turn into Mn(III) (Fig.~\ref{fig7}d).

The possibility of an EC adsorbed on a surface Mn(III) ion is also considered.
Upon optimizing the Kohn-Sham wavefunctions and the geometry, the
Mn(III) coordinated to the EC carbonyl oxygen reverts to an Mn(IV),
with the extra electron migrating to another surface Mn(IV), converting
that Mn to a Mn(III) (not shown).  This suggests that the charge states
of surface Mn ions strongly depend on whether molecules are coordinated
to them, at least on this (100) surface.

We have also considered a larger, Li$_8$Mn$_{24}$O$_{48}$ slab, of
the same lateral dimension as the Li$_6$Mn$_{20}$O$_{40}$ model studied
in most of this work.  The tetrahedral site occupancies are determined
by calculating the energies of all symmetry-distinct Li-configurations.
This model features 8~Mn(III) ions, and in principle all 8~surface Mn
could have been Mn(III).  Instead, 2~of~the~8 surface Mn remain Mn(IV),
presumably because the surface layers are deficient in Li$^+$.  

\section*{Appendix B: Hydroxyl and proton on (100) spinel MnO$2$ surface}
\label{proton}

Interactions between H$_2$O and Li$_x$Mn$_2$O$_4$ surfaces are pertinent to
how trace H$_2$O in the atmosphere or electrolyte affects Mn$_2$O$_4$
stability.  (Water may also emerge from electrolyte decomposition.)  Acids are
known to degrade spinel oxide electrodes,\cite{thackeray,acid_bulk} and water
can be a source of H$^+$.  In this appendix, we consider the effect of
decorating and functionalizing Li$_x$Mn$_2$O$_4$ (100) surfaces with H$_2$O,
H$^+$, and OH$^-$ groups.

\begin{figure}
%\centerline{\hbox{ (a) \epsfxsize=1.50in \epsfbox{fig6a.ps}
%                   (b) \epsfxsize=1.50in \epsfbox{fig6b.ps} }}
%\centerline{\hbox{ (c) \epsfxsize=1.50in \epsfbox{fig6c.ps}
%                   (d) \epsfxsize=1.50in \epsfbox{fig6d.ps} }}
\centerline{\hbox{ (a) \epsfxsize=2.25in \epsfbox{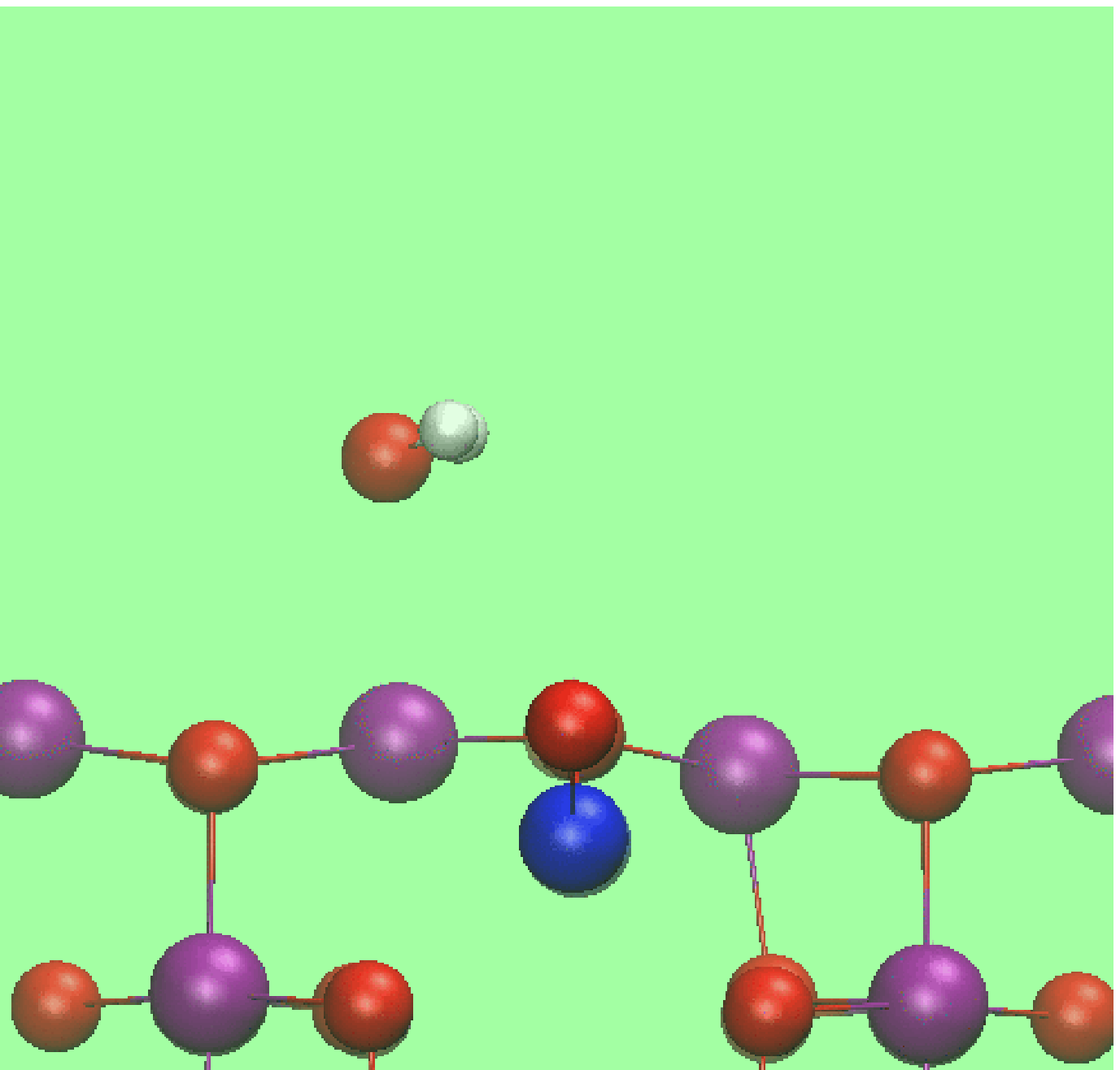}
                   (b) \epsfxsize=2.25in \epsfbox{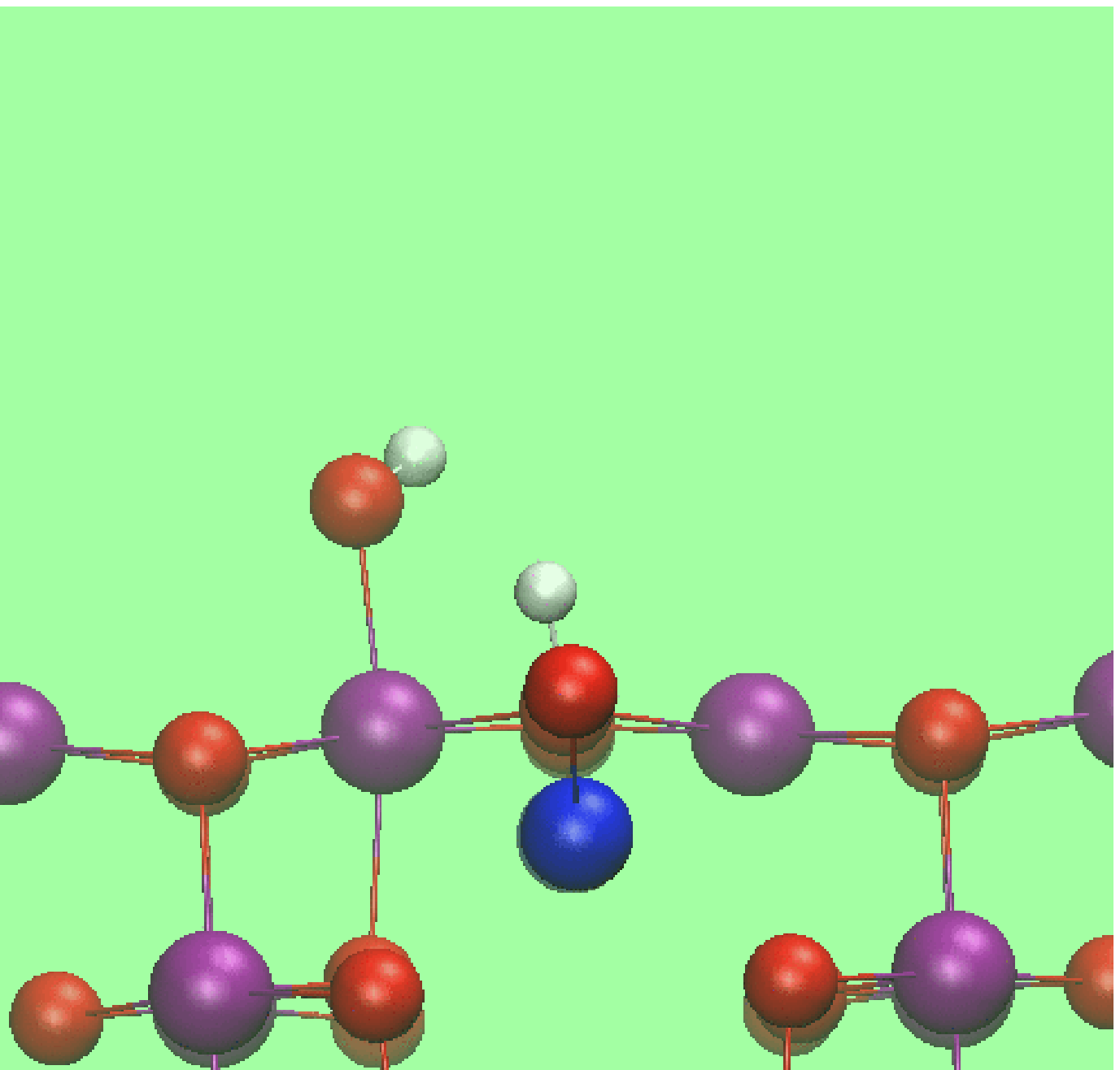} }}
\centerline{\hbox{ (c) \epsfxsize=2.25in \epsfbox{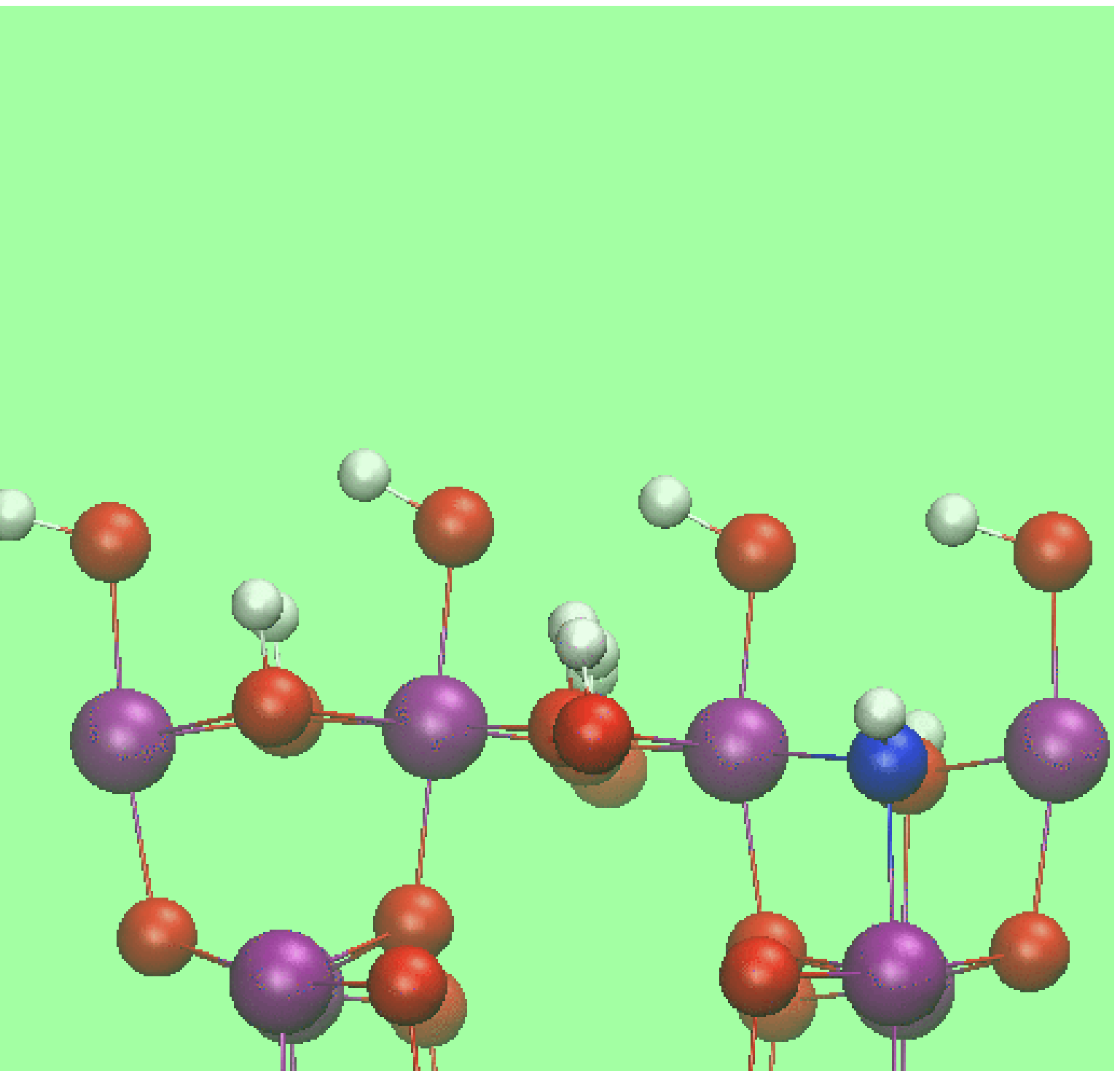}
                   (d) \epsfxsize=2.25in \epsfbox{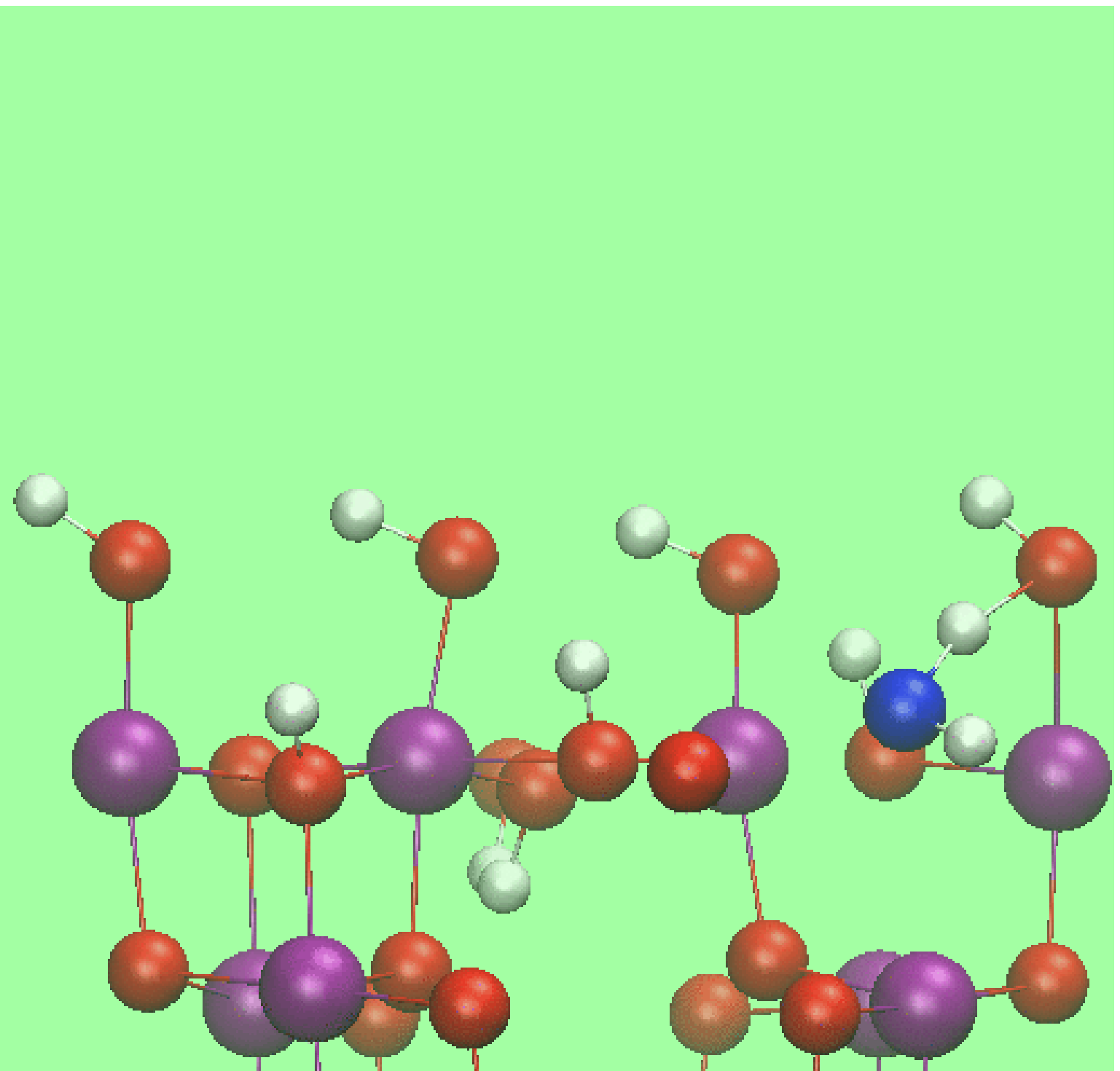} }}
\caption[]
{\label{fig6} \noindent
(a)\&(b): Intact and dissociated H$_2$O molecule on MnO$_2$ surface. 
In panel (a), one of the water protons is obscured by the O~atom.
(c)\&(d): The initial and final hydroxylated and protonated MnO$_2$ 
(100) surface configurations along an AIMD trajectory, respectively.  In
panel (d), one OH group on the surface has abstracted a proton from another
hydroxyl group and has turned into a H$_2$O molecule; that
oxygen is colored deep blue.
}
\end{figure}

Figure~\ref{fig6}a depicts H$_2$O physisorption on the (100) surface
at the $x=0.6$ stochiometry.  It exhibits a binding energy of 0.452~eV,
comparable to EC physisorption.  Breaking a O-H
bond in water and displacing the proton to a neighboring surface O~ion
to form two new hydroxyl groups lead to a 1.027~eV binding energy after
performing geometry optimization (Fig.~\ref{fig6}b).  Thus dissociative
chemisorption is more favorable than physisorption by 0.575~eV.  However,
even this chemisorbed configuration may be metastable compared to one where
the H$^+$ has penetrated into the subsurface layer (see below).

Next, we consider the effect of multiple hydroxyl groups on
the surface.  We turn to Mn$_2$O$_4$, without Li$^+$ intercalation,
and decorate all surface O~ions with H atoms and all surface Mn ions
with OH groups.  The geometry is relaxed to a local energy minimum
(Fig.~\ref{fig6}c).  The first type of hydroxyl groups, originating
from protonation of surface oxygen atoms, point outward or
parallel to the oxide surface.  The second type, hydroxyl groups formed
by adding OH pendants to Mn, always point out of the plane.  As there are
more added H~than~OH, the charge-neutral slab now contains some Mn(III) ions.  

Starting from this optimized geometry, a 11~ps AIMD trajectory is conducted
at T=450~K.  The final configuration is shown in Fig.~\ref{fig6}d.
Some type~1 hydroxyl groups now point inwards, with their H$^+$ occupying
sites that Li$^+$ normally reside in Li$_x$Mn$_2$O$_4$.  This is
reasonable because the interior sites bracketed by O$^{2-}$ ions should
be electrostatically more favorable for cations and protons.  In one
instance, a type~1 hydroxyl group has abstracted a H$^+$ from a neigboring
hydroxyl group to form a H$_2$O.  The oxygen atom of that water molecule
coordinates to a surface Mn~ion 2.3~\AA\, away in this snapshot, instead
of the initial Mn-O ionic bonding distance ($\sim 1.9$~\AA).  That Mn
ion now has only 4 ionic bonds in addition to coordination to the H$_2$O
molecule, and should be more susceptable to chemical reactions and
dissolution.  These trends are consistent with the experimental observations
that protons accelerate decomposition of spinel
Li$_x$Mn$_2$O$_4$.\cite{thackeray,pasquier,jang1997} Despite the low
temperature compared to experimental annealing conditions during
synthesis,\cite{huang2001} evidence of H$_2$O loss is already observed.

Type~2 hydroxyl groups appear stable over the course of the AIMD trajectory.
As surface O~atoms dissolves after H$^+$ attacks, however, even Mn~ions attached
to type 2 hydroxyls may become undercoordinated, exposed, and susceptable
to attack by other H$_2$O moleules or the organic electrolyte.

\newpage

\newpage
 
\end{document}